\documentclass[sn-mathphys-num]{sn-jnl}

\usepackage{graphicx}%
\usepackage{multirow}%
\usepackage{amsmath,amssymb,amsfonts}%
\usepackage{amsthm}%
\usepackage{mathrsfs}%
\usepackage[title]{appendix}%
\usepackage{xcolor}%
\usepackage{textcomp}%
\usepackage{manyfoot}%
\usepackage{booktabs}%
\usepackage{algorithm}%
\usepackage{algorithmicx}%
\usepackage{algpseudocode}%
\usepackage{listings}%

\usepackage{times}

\usepackage{url}
\usepackage{hyperref}
\usepackage{amsmath} 
\usepackage{latexsym}
\usepackage{bm} 
\usepackage{graphics} 
\usepackage{arydshln} 
\usepackage{graphicx} 
\usepackage{float} 
\usepackage{threeparttable} 
\usepackage{pifont} 
\usepackage{xcolor}
\newcommand{\cmark}{\ding{51}}%
\newcommand{ \xmark }{\ding{55}}%
\newcommand{\dotsim}{\mathrel{\dot\sim}}
\renewcommand{\log}{\text{log}}

\theoremstyle{thmstyleone}%
\theoremstyle{thmstyletwo}%

\theoremstyle{thmstylethree}%

\begin{document}

\title[Robust Estimation for Missing Exposure]{Causal Inference on Missing Exposure via Robust Estimation}


\author[1]{Yuliang Shi }\email{yuliang.shi@uwaterloo.ca}

\author[1]{Yeying Zhu}\email{yeying.zhu@uwaterloo.ca}

\author[1]{Joel A. Dubin}\email{jdubin@uwaterloo.ca}

\affil*[1]{\orgdiv{Department of Statistics and Actuarial Science}, \orgname{University of Waterloo}, \orgaddress{\street{200 University Ave W}, \city{Waterloo}, \postcode{N2L 3G1}, \state{Ontario}, \country{Canada}}}


\abstract{How to deal with missing data in observational studies is a common concern for causal inference. When the covariates are missing at random (MAR), multiple approaches have been provided to help solve the issue. However, if the exposure is MAR, few approaches are available and careful adjustments on both missingness and confounding issues are required to ensure a consistent estimate of the true causal effect on the response. In this article, a new inverse probability weighting (IPW) estimator based on weighted estimating equations (WEE) is proposed to incorporate weights from both the missingness and propensity score (PS) models, which can reduce the joint effect of extreme weights in finite samples. Additionally, we develop a triple robust (TR) estimator via WEE to further protect against the misspecification of the missingness model. The asymptotic properties of WEE estimators are proved using properties of estimating equations. Based on the simulation studies,  WEE methods outperform others including imputation-based approaches in terms of bias and variability. Finally, an application study is conducted to identify the causal effect of the presence of cardiovascular disease on mortality for COVID-19 patients.}




\maketitle

\section{Introduction}\label{sec1}
In biostatistical or epidemiological studies, it is often important for researchers to estimate the causal effect of an exposure of interest on the primary outcome. However, there may exist multiple confounders that jointly affect the exposure and the outcome. A common tool to adjust for confounding issues is the propensity score (PS), which is defined as the probability of receiving treatment given the covariates  \cite{rosenbaum}.  However,  estimation of the propensity score as well as the causal effect is challenging if there exists missing data on the exposure of interest. In the literature, few approaches have been proposed to deal with this issue, under the missing at random (MAR) assumption \cite{Rubin}. In this paper, we will focus on the problem of estimating causal effects when we need to adjust for both missing and confounding issues. 


In our motivating example, we aim to investigate the causal effect of an exposure variable, i.e. presence of cardiovascular disease (CVD), on mortality, from an observational cohort of COVID-19 patients in Brazil. However, in this observational study, CVD status was not fully observed because the health status of patients was not fully recorded during the pandemic. In addition, some clinical variables, such as age, sex, diabetes, etc, may affect both CVD status and death, so the confounding issues should also be carefully dealt with \cite{leon2015diabetes}. In other words, one can view this problem as a missing data problem of two layers, involving (1) the exposure status being missing for some individuals and (2) half of the potential outcomes are not observed. That motivates us to develop a robust method to adjust for both missing and confounding issues and protect against the misspecification of models at the same time.

One of the common approaches to dealing with missing data is via imputation. Single imputation (SI) is easy to conduct based on the observed data \cite{donders2006gentle}. However, researchers have suggested that conducting the imputation only once is not always reliable, and conducting the imputation multiple times can reduce the bias and the variance of the estimators \cite{Buuren}. \citet{buurenmice} propose the multiple imputations chained equations (MICE) approach, and a well-known R package ``mice" is available to implement this method \cite{buuren2011}. It utilizes a Gibbs sampler for random sampling, and Rubin's rules can be utilized to combine results from all imputed datasets \cite{rubin2004multiple}. However, parametric imputation-based methods must rely on a correctly specified imputation model. If researchers misspecify the imputation model, even though both PS and outcome models are correct, the estimation of causal effects is biased. Moreover, \citet{Seaman} also state that, for imputation-based methods, there may exist the extrapolation issue from the observed data to the missing data, and there is little scope to assess whether the explicit extrapolation can be justified due to the missing values. 

In contrast,  inverse probability weighting-based (IPW-based) methods have also been widely investigated because they do not necessarily require the imputation model to be correct, and they may not suffer from the extrapolation issue \cite{neyman1923application}. In the previous literature, \citet{williamson2012doubly} propose a two-layer double robust (DR) estimator to adjust for the missing and confounding issues based on the missingness, the imputation, PS, and the outcome models.  \citet{williamson2012doubly} state that a two-layer DR estimator requires the ``one of two models" condition, i.e. one correct model from either the missingness or the imputation model and another correct model from either PS or the outcome model. Later on, \citet{zhang} propose a triple robust (TR) estimator, which requires the ``two of three models" condition to achieve consistency, i.e. at least two correct models from the missingness, PS, and outcome models. Besides that, \citet{kennedy2020efficient} propose the nonparametric estimators with the efficient bounds using the efficient influence function, which requires PS model to be correct and either the imputation or the missingness model is also correct. 



However, some limitations exist in previous studies. First, these articles \cite{williamson2012doubly,zhang,kennedy2020efficient} do not describe the effect of extreme weights on the estimated results in finite samples. Based on our simulation studies (see Section \ref{sec_simulation}), we observe that when some extreme weights occur, two-layer DR estimator and previous TR estimator result in larger bias and standard errors in finite samples. Although truncation and stabilization for extreme weights have been previously proposed in the literature \cite{lee2011weight,zhou2020propensity}, the cut-off values for IPW trimming are usually arbitrarily decided, and when those extreme weights have quite larger effects on the estimated values than other weights, stabilization for extreme weights may not be helpful. In addition, the robust properties can be improved to protect against the misspecification of the missingness or imputation model. For example, \citet{williamson2012doubly} do not mention how to solve the problem when both missingness and imputation models are wrong.  In comparison, \citet{zhang} require both PS and outcome models to be correct when the missingness model is wrong. The nonparametric estimators from \citet{kennedy2020efficient} require PS model to be correct for consistency, which is restrictive in application. Finally, the inference results should be further investigated for the robust estimators, because the simulation studies from \citet{williamson2012doubly} show very high coverage rates of confidence intervals (CI) when the outcome model is wrong without providing detailed reason in the article.

These limitations motivate us to propose a new IPW method based on weighted estimating equations (WEE), called the ``IPW-WEE" estimator. The idea is to construct IPW-WEE in a ``two-step procedure" to separate the joint effects of inverse weights from the missingness and PS models, which is inspired by the DR estimator using weighted least squares (WLS) \cite{kang2007demystifying,robins2007comment}. In addition, we further develop TR estimators to protect against the misspecification of either the missingness or the imputation model. TR-WEE is also more resistant to extreme weights than the previous estimators in finite samples. If we classify the missingness and imputation models as the first group, PS model as the second group, and the outcome model as the third group, TR-WEE only requires ``two out of three groups of models to be correct" to achieve consistency and asymptotic normality, called ``TR group properties". Our simulation studies also show that the coverage rates of IPW-WEE and TR-WEE are close to 95\% with the bootstrap percentile approach as the valid inference results. Finally, the performance of IPW and TR estimators is also evaluated and compared with imputation-based methods in both simulation and application studies.

Some articles have proposed estimators that are ``triple robust" or ``multiply robust". To clarify the difference, \citet{zhang} state their TR properties as ``two models out of three models to be correct" because they only consider the missingness, PS, and outcome models without the imputation model. In comparison, \citet{williamson2012doubly} consider the imputation model as one of the required models, but they do not use the word ``triple robust". Instead, they separate the models into two groups, including either (1) the missingness and imputation models or (2) PS and outcome models, and claim that the estimator is consistent if ``one of two models plus another one of two models to be correct". Besides that, multiply robust estimators have been proposed in the literature using the empirical likelihood which requires any one of multiple models to be correctly specified \cite{han2014multiply}, but the estimators are usually constructed when the outcome is missing, which is a different scenario from MAR on the exposure variable. In contrast, TR group properties is defined as ``two groups out of three groups of models to be correct" in this article.

To make results available for other users, we have also developed an R package for TR estimators to solve the problem of the missing exposure, called \href{https://github.com/yuliang-shi/trme}{``trme"} uploaded on \href{https://github.com/yuliang-shi}{GitHub website}. The package can provide reliable point estimation of the causal effect and statistical inference to deal with the missing exposure and confounding issues in real applications.  

The remainder of this article is organized as follows. In Section 2, we describe the goal, the notation, and the models. Section 3 provides the theory of IPW-based estimators. In Section 4, we discuss TR theory and the connection among different estimators. Section 5 provides simulation studies to compare the performance of different methods. In Section 6, an analysis of the motivating application is conducted to identify the causal effect of cardiovascular disease on mortality for a cohort of COVID-19 patients. The article ends with a discussion in Section 7.

\section{Notations and Models}
\label{sec_notation}

\subsection{Notations and Causal Framework}

In this section, we present the notations and the models for our proposed methods. The counterfactual causal framework \cite{rubin1974estimating}) denotes $(Y_i^1, Y_i^0), i=1,\dots,n,$ as the potential outcomes if individual $i$ were exposed or unexposed (or equivalently, the treated or untreated), respectively. Let $\bm{X}_i=(1,X_{i1},...,X_{i,p})^T$  denote a $(p+1) \times 1$ vector of an intercept term and  $p$ covariates for individual $i$. We further denote $\mathbf{X}=(\bm{X_{1}},\bm{X_2},...,\bm{X_{n}})^T$ as the $n \times (p+1)$ design matrix. Let $A_i$ denote the exposure of interest, $Y_i$ denote the binary outcome, and $R_i=I\{A_i \text{ is missing}\}$ denote the indicator of whether the exposure value is missing or not for individual $i$. In this study, we only consider the exposure variable to be missing, so all other variables are assumed to be completely observed. However, the idea about reducing the joint effect of extreme weights through WEE can be extended to the other situation, such as MAR on both exposure and outcome if we modify WEE estimators and MAR assumption.  

Notice that for each individual $i$, $Y_i^1$ and $Y_i^0$ cannot be observed at the same time; instead, we observe the actual exposure $A_i$ and corresponding outcome $Y_i$. The relationship between the observed outcome and the potential outcomes is $Y_i=A_iY_i^1+(1-A_i)Y_i^0$. The observed dataset can be written as $(\bm{X}_i,A_i,Y_i,R_i), i=1, \dots, n$, and the true propensity score (PS) is defined as $P_{A_i}(\bm{X}_i)=P(A_i=1|\bm{X}_i)$ for individual $i=1,\dots,n$. Since PS is a balancing score, i.e. $\bm{X} \bot A|P_A (\bm{X})$, all measured confounders between the exposure and non-exposure groups are balanced given PS values \cite{rubin1974estimating}. This is the key reason why we can apply the inverse weights of the propensity score to estimate the causal effect in the next section.

Our main interest is to identify the causal effect of the exposure on the outcome when the exposure is MAR. Motivated by our application study, we focus on a binary outcome and denote $\tau_1=E(Y^1)=P(Y^1=1)$ and $\tau_0=E(Y^0)=P(Y^0=1)$, so the true causal effect $\tau$ is defined as the odds ratio (OR) of the potential outcomes if exposed versus not exposed:
\begin{equation}
\begin{aligned}
\label{truecausal}
\tau=\frac{\tau_1/(1-\tau_1)}{\tau_0/(1-\tau_0)}=\frac{P(Y^1=1)/P(Y^1=0)}{P(Y^0=1)/P(Y^0=0)}.
\end{aligned}
\end{equation}
Certainly, other estimands such as the average causal effect or risk ratio can be considered and our proposed estimators need to be modified accordingly. 

To estimate $\tau$ as the causal odds ratio in observational studies, three assumptions are required in this study, described as follows:
\begin{enumerate}
    \item Strongly ignorable treatment assignment assumption (SITA): $(Y^1, Y^0) \bot A|\bm{X}$.
   \item Positivity assumption:  $0<P(A=1|\bm{X}=\bm{x})<1, \text{ and } 0<P(R=1|\bm{X}=\bm{x},\bm{Y}=\bm{y})<1 \text{ for all possible }  \bm{x, y}.$

       
    \item MAR assumption:  $R \bot A|(\bm{X},Y)$
\end{enumerate}

\begin{figure}[ht]
\includegraphics[width=\textwidth,height=2.2cm]{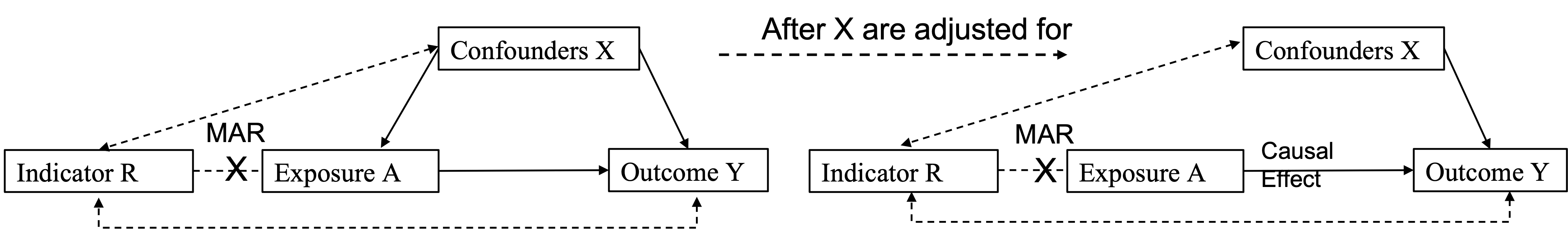}
\caption{Causal diagram before and after the confounders are adjusted for. The black solid arrows refer to the causal relationship among confounders, exposure, and outcome. The double-sided dash arrows refer to the associational relationship among the missing indicator, the outcome, and the covariates. Due to the MAR assumption, the dashed short line refers to no association between the missing indicator and the exposure given covariates and the outcome.}
\label{fig:causal_diag}
\end{figure}

SITA assumption is also known as the assumption of no unmeasured confounder, which cannot be tested using the observed data.  The positivity assumption guarantees that the propensity score and the missingness probability are non-zero, which can be checked on observed data when other covariates are fully observed. The MAR assumption describes the missingness mechanism for the exposure variable, which assumes the missing indicator is conditionally independent of the treatment itself given all other observed variables in the dataset. However, MAR assumption cannot be tested either because we do not know the true missing values. The basic causal framework can be described in Figure \ref{fig:causal_diag}.  

\subsection{Identification}

As we have previously mentioned, estimating $\tau$ requires us to account for both missing and confounding issues, which can be viewed as a ``two-layer" missing data problem. Assumptions (1)-(3) ensures the identification for $\tau_1$ on the observed data, written as
\begin{equation}
\begin{aligned}
\label{identify}
E \left[ \frac{1-R}{1-E(R|\bm{X},Y)}\frac{AY}{E(A|\bm{X})} \right]
&= E \left\{ E\left[ \frac{1-R}{1-E(R|\bm{X},Y)} \Bigg|\bm{X},Y \right] E\left[ \frac{AY}{E(A|\bm{X})}\Bigg|\bm{X},Y \right] \right\} \\
&=E \left\{ E \left[ \frac{AY}{P(A=1|\bm{X})} \Bigg|\bm{X},Y^1 \right]  \right\} \\
&=E \left[ \frac{E(A|\bm{X})Y^1}{P(A=1|\bm{X})}  \right]
=E(Y^1),
\end{aligned}    
\end{equation}
where the first equality is due to the property of the double expectation on $(\bm{X},Y)$ and MAR assumption. The second equality holds due to the double expectation on $(\bm{X},Y^1)$. The third equality holds because of SITA and the definition of $Y=AY^1+(1-A)Y^0$. The last equality holds because we know $E(A|\bm{X})=P(A=1|\bm{X})$. Similarly, we know that $\tau_0$ can also be identified as $E \left[ \frac{1-R}{1-E(R|\bm{X},Y)}\frac{(1-A)Y}{1-E(A|\bm{X})} \right]=E(Y^0)$. Here, both $\tau_0, \tau_1$ can be expressed as the function of $(R, A,\bm{X},Y)$, which is the observed dataset, so the identification for $\tau$ can be guaranteed.

\subsection{Models}

In this article, we mainly consider the parametric framework with four regression models, which are classified into three groups: 
\begin{itemize}
    \item (1) the missing/imputing group includes the missingness model: $P_{R}(\bm{\gamma})=P(R=1|\bm{X},Y;\bm{\gamma})=\text{expit}(\bm{X}^T\bm{\gamma}+Y\gamma_y)$, and the imputation model: $P(A=1|\bm{X}, Y;\bm{\delta})=\text{expit}(\bm{X}^T\bm{\delta}+Y\delta_y)$;
    \item (2) the PS model as the second group: $P_{A}(\bm{\alpha})=P(A=1|\bm{X}; \bm{\alpha})=\text{expit}(\bm{X}^T\bm{\alpha})$; 
    \item  (3) the outcome model as the third group and its model form depends on the type of response. Following the motivated example, we mainly consider the binary outcome, so the outcome model is written as: $P_Y(\bm{\beta})=E(Y|\bm{X}, A; \bm{\beta})=\text{expit}(\bm{X}^T\bm{\beta}+A\beta_A)$.
\end{itemize}
Here, $(\bm{\gamma},\bm{\delta},\bm{\alpha},  \bm{\beta}, \gamma_y,\delta_y,\beta_A)$ refers to the unknown parameters that need to be estimated. For simplicity, we will just use $(\bm{\gamma},\bm{\delta},\bm{\alpha},  \bm{\beta})$ to denote all unknown parameters and their true values are denoted as $(\bm{\gamma^*},\bm{\delta^*}, \bm{\alpha^*}, \bm{\beta^*})$. 
 We define $\bm{\beta_1^*}$ and $\bm{\beta_0^*}$ as the true coefficients of the outcome model for the exposure and non-exposure group, respectively. If the model is wrongly specified, we denote the incorrect values of the coefficients as $(\bm{\widetilde{\gamma}},\bm{\widetilde{\alpha}},\bm{\widetilde{\beta}},\bm{\widetilde{\delta}})$. 
 
 Even though the study focuses on a binary outcome, the theories are developed in general scenarios, so the application can be easily expanded to other cases, such as a continuous or count outcome. The major component that needs to be changed is the form of the outcome model in the estimating equation, shown in Section \ref{sec:review_alpha_wla}.



Notice that the coefficients for the outcome models, i.e. $\bm{\beta}'s$, are the nuisance parameters, which are not considered as the main interest, and they just refer to the associational instead of the causal relationship between the exposure and the outcome. However,  $\bm{\hat{\beta}}$ is used as the plugged-in values for the estimating equation of IPW or TR estimators. In addition, all proposed methods in the following sections are discussed under the general semiparametric framework \cite{van2000asymptotic}.

\section{IPW Estimators}

\subsection{Review of Two Estimators for Model Parameters}
\label{sec:review_alpha_wla}

Before we introduce estimators for $\tau$, 
we first talk about how to estimate the coefficients in the PS model by adjusting for the missingness in the exposure under MAR assumption. More specifically, we aim to apply an IPW-based method to estimate $\bm{\alpha}$ after adjusting for the missingness in PS model. We call  it the weighted likelihood approach (WLA) and  denote the estimated value of $\bm{\alpha}$ as $\bm{\hat{\alpha}}^{(WLA)}$, such that $\bm{\hat{\alpha}}^{(WLA)}$ solves the following estimating equation:
\begin{equation}
\begin{aligned}
\label{alpha_WLA}
S({\bm{\alpha}}^{(WLA)})=\sum_{i=1}^n \frac{1-R_i}{1-P_{R_i}(\bm{\hat{\gamma}})} U_i(A_i,\bm{X}_i;\bm{\alpha})=0.
\end{aligned}    
\end{equation}
Here, $U_i(A_i,\bm{X}_i; \bm{\alpha})=\bm{X}_i[A_i-E(A_i|\bm{X}_i;\bm{\alpha})]$ is  $(p+1) \times 1$ vector of the original score function from PS model for individual $i$, and $\bm{\hat{\gamma}}$ is an estimate of $\bm{\gamma}$. If the missingness and PS models are correct, we know $\bm{\hat{\alpha}}^{(WLA)} \xrightarrow[]{\text{p}} \bm{\alpha^*}$ and its asymptotic normality holds as $n \to \infty$ based on the proof for $\hat{\bm{\alpha}}^{(EE)}$ in S.1 of supplementary material. 

Similarly, we also need to adjust for the missingness on the outcome model to consistently estimate the coefficients $\bm{\beta}$. We assume the missingness model is  correct, and apply WLA to obtain $\bm{\hat{\beta}}^{(WLA)}$, which solves the following estimating equation:
\begin{equation}
\begin{aligned}
\label{beta_WLA}
S({\bm{\beta}}^{(WLA)})=\sum_{i=1}^n \frac{1-R_i}{1-P_{R_i}(\bm{\hat{\gamma}})} V_{i}(Y_i,A_i, \bm{X}_i; \bm{\beta}) =0,
\end{aligned}    
\end{equation}
where $V_{i}(Y_i,A_i, \bm{X}_i; \bm{\beta})=\bm{X}_i[Y_i-E(Y_i|A_i,\bm{X}_i;\bm{\beta})]$ is the $(p+2) \times 1$ vector of the original score function from the binary outcome model with the covariates and the exposure. If some researchers aim to study for the continuous outcome, the score function can be changed into  $V_{i}(Y_i,A_i, \bm{X}_i; \bm{\beta})=\bm{X}_i(Y_i-\bm{X}_i^T\bm{\beta})/\sigma^2$. If both missingness and outcome models are correct, we know $\bm{\hat{\beta}}^{(WLA)} \xrightarrow[]{\text{p}} \bm{\beta^*}$ and its asymptotic normality holds as $n \to \infty$. Notice that $(\bm{\hat{\alpha}}^{(WLA)}, \bm{\hat{\beta}}^{(WLA)})$ are not considered as the main interest, instead, they are just plugged-in values for unknown $(\bm{\alpha,\beta})$ in the IPW estimators in the next subsection.

\subsection{Review of Two IPW Estimators}

In this subsection, we will review two estimators for $\tau$ using IPW method, including IPW-IPW and IPW-DR, where the first part of the name refers to the method used for missing data and the second part means the method used for the confounding issue \cite{zhang}.  In the following discussion, for all IPW estimators, we require the missingness model to be correct, otherwise, the estimated causal effect is biased. We will release this requirement when we develop TR estimators in  Section \ref{sec_TR}.  We will mainly discuss the estimator for $\tau_1$, because the estimator for $\tau_0$ is straightforward to develop after replacing $A_i$ by $1-A_i$.

The idea of IPW-IPW is to directly apply two inverse weights on the observed data to directly adjust for the missingness and confounding issues, which requires both missingness and PS models to be correct. Based on the identification formula of Equation (\ref{identify}),  the empirical IPW-IPW estimator, denoted as $\hat{\tau}_1^{(IPW-IPW)}$, can be written as:
\begin{equation}
\begin{aligned}
\label{IPW-IPW}
\hat{\tau}_1^{(IPW-IPW)}=\frac{1}{n}\sum_{i=1}^n \frac{1-R_i}{1-P_{R_i}(\bm{\hat{\gamma}})}\frac{A_iY_i}{P_{A_i}(\hat{\bm{\alpha}}^{(WLA)})},
\end{aligned}    
\end{equation}
 where $\bm{\hat{\gamma}}$ is estimated via the missingness model defined in Section \ref{sec_notation}.  $P_{A_i}(\hat{\bm{\alpha}}^{(WLA)})$ is fitted PS values after plugging in $\hat{\bm{\alpha}}^{(WLA)}$. When both missingness and PS models are correct,  we can prove $\hat{\tau}_1^{(IPW-IPW)} \xrightarrow[]{\text{p}} \tau_1$ and its asymptotic normality hold as $n \to \infty$. Then, $\hat{\tau}^{(IPW-IPW)}= \frac{\hat{\tau}_1^{(IPW-IPW)}/(1-\hat{\tau}_1^{(IPW-IPW)})}{\hat{\tau}_0^{(IPW-IPW)}/(1-\hat{\tau}_0^{(IPW-IPW)})} \xrightarrow[]{\text{p}} \tau$.

Since IPW-IPW estimator requires both models to be correct, which is too restrictive, the second IPW-DR estimator is developed to protect against the misspecification of either PS or the outcome model,  written as:
\begin{equation}
\begin{aligned}
\label{IPW-DR}
\hat{\tau}_1^{(IPW-DR)}
&=\frac{1}{n}\sum_{i=1}^{n}  \frac{1-R_i}{1-P_{R_i}(\bm{\hat{\gamma}})} \left\{ \frac{A_i}{P_{A_i}(\bm{\hat{\alpha}}^{(WLA)})}\Delta(Y_i,\bm{X}_i; \hat{\bm{\beta}}^{(WLA)}_1) +E \left[Y_i|A_i=1,\bm{X}_i,\hat{\bm{\beta}}_1^{(WLA)} \right] \right\}
\end{aligned}    
\end{equation}
where $\bm{\hat{\gamma}}$ are estimated via the missingness model and $\Delta(Y_i,\bm{X}_i; \hat{\bm{\beta}}_1)=Y_i-E(Y_i|A_i=1,\bm{X}_i;\hat{\bm{\beta}}_1)$ is denoted as the residual term in Equation (\ref{IPW-DR}). If the missingness model is required to be correct and either PS or the outcome model is also correct, we know $\hat{\tau}_1^{(IPW-DR)} \xrightarrow[]{\text{p}} \tau_1$ and its asymptotic normality hold as $n \to \infty$ \cite{zhang}. 
In addition, we can obtain an estimator for $\tau_0$ after replacing the exposure group with the non-exposure group, and $\hat{\tau}^{(IPW-DR)}= \frac{\hat{\tau}_1^{(IPW-DR)}/(1-\hat{\tau}_1^{(IPW-DR)})}{\hat{\tau}_0^{(IPW-DR)}/(1-\hat{\tau}_0^{(IPW-DR)})} \xrightarrow[]{\text{p}} \tau$ as $n \to \infty$.

\subsection{IPW-WEE Estimator}
\label{sec:ipw_wee_theory}

The drawback of both IPW-IPW and IPW-DR estimators is that they involve the product of the estimated missingness probability and the propensity score in the denominator, and the weights may become extremely large when the estimated probability is close to one.  That motivates us to develop a so-called weighted estimating equation (WEE) approach for IPW-based estimators, which is expected to be more resistant to extreme weights.

IPW-WEE estimator is proposed in two steps to separate the joint effect of extreme weights on $\hat{\tau}_1$ from the missingness and PS models. In the first step, we revise Equation (\ref{beta_DR_Miss}) to apply both inverse weights of the missingness and PS values to estimate $\bm{\beta}$ and denote the new estimate of the outcome coefficients as $\hat{\bm{\beta}}_1^{(IPW-WEE)}$ (where the subscript ``1" refers to the exposure group) by solving the following equation: 
\begin{equation}
\begin{aligned}
\small
\label{beta,IPW-WEE}
S(\bm{\beta}_1^{(IPW-WEE)})
&= \sum_{i=1}^{n} \bm{X}_i  \frac{1-R_i}{1-P_{R_i}(\bm{\hat{\gamma}})} \frac{A_i}{P_{A_i}(\bm{\hat{\alpha}}^{(WLA)})}[Y_i-E(Y_i|A_i=1,\bm{X}_i;\bm{\beta}_1)]=0
\end{aligned}
\normalsize
\end{equation}
where $E(Y_i|A_i=1,\bm{X}_i;\bm{\beta}_1)=\text{expit}(\bm{X}_i^T \bm{\beta}_1)$ is the outcome model for the exposure group with unknown $(p+1) \times 1$ vector of $\bm{\beta}_1$ to be solved. Here, $\bm{X}_i$ requires to include one intercept and $p$ covariates. Since the Fisher information matrix is positive definite, estimating equation (\ref{beta,IPW-WEE})  leads to a unique solution for $\bm{\beta}_1$.  If both missingness and outcome models are correct, we prove that $\hat{\bm{\beta}}_1^{(IPW-WEE)} \xrightarrow[]{\text{p}} \bm{\beta}_1^*$ and the asymptotic normality holds as $n \to \infty$. The detailed proof of the asymptotic properties for $\hat{\bm{\beta}}_1^{(IPW-WEE)}$ is provided in S.2 of the supplementary material. If researchers consider the continuous outcome, we can simply modify Equation (\ref{beta,IPW-WEE}) with the linear regression model, i.e. $E(Y_i|A_i=1,\bm{X}_i;\bm{\beta}_1)=\bm{X}_i^T \bm{\beta}_1$.

In the second step, we estimate $\tau_1$ by:
\begin{equation}
\begin{aligned}
\label{IPW-WEE}
\hat{\tau}_1^{(IPW-WEE)}=
&\frac{1}{n} \sum_{i=1}^n \frac{1-R_i}{1-P_{R_i}({\bm{\hat{\gamma}}})} E \left[Y_i \Big|A_i=1,\bm{X}_i,\hat{\bm{\beta}}_1^{(IPW-WEE)} \right],
\end{aligned}
\end{equation}
where $E \left[Y_i|A_i=1,\bm{X}_i,\hat{\bm{\beta}}_1^{(IPW-WEE)} \right]=\text{expit}(\bm{X}_i^T \hat{\bm{\beta}}_1^{(IPW-WEE)})$ is the predicted response for each individual in the exposure group after plugging in $\hat{\bm{\beta}}_1^{(IPW-WEE)}$. Notice that $\hat{\tau}_1^{(IPW-WEE)}$ in Equation (\ref{IPW-WEE}) can be viewed as the weighted average on fitted responses for the exposure group, which only involves inverse weights of the missingness without any weights from PS model.

If the missingness model is correct and either PS or the outcome model is correct, we prove that $\hat{\tau}_1^{(IPW-WEE)} \xrightarrow[]{\text{p}} \tau_1$ and its asymptotic normality holds as $n \to \infty$. Similarly, $\hat{\tau}_0^{(IPW-WEE)}$ is a IPW-WEE estimator for $\tau_0$ after we replace $A_i$ with $1-A_i$ and we can also prove $\hat{\tau}_0^{(IPW-WEE)} \xrightarrow[]{\text{p}} \tau_0$. Then, IPW-WEE estimate for $\tau$ will be consistent to the true causal effect as $n \to  \infty$, written as:
\begin{equation}
\begin{aligned}
\hat{\tau}^{(IPW-WEE)}= \frac{\hat{\tau}_1^{(IPW-WEE)}/(1-\hat{\tau}_1^{(IPW-WEE)})}{\hat{\tau}_0^{(IPW-WEE)}/(1-\hat{\tau}_0^{(IPW-WEE)})} \xrightarrow[]{\text{p}} \tau.
\end{aligned}    
\end{equation}
Detailed proofs for the connection between IPW-WEE and IPW-DR are also provided in S.2 of the supplementary material. In addition, we also prove the asymptotic normality for $\hat{\tau}$ provided in S.4 of the supplementary material based on the properties of estimating equations, Taylor expansion, and the delta method.  

In summary, we provide an algorithm to conduct IPW-WEE estimator in four stages:

\begin{enumerate}
    \item Fit the missingness model and obtain $\bm{\hat{\gamma}}$ via $R\sim \bm{X}+Y$ as the plugged-in values in Step 3 (a);
    \item Solve $\hat{\bm{\alpha}}^{(WLA)}$ from Equation (\ref{alpha_WLA}), which contains PS model via $A \sim \bm{X}$, as the plugged-in values in Step 3 (a).
    \item Conduct the ``two-step procedure"  on $\hat{\tau}_1$:
        \begin{enumerate}
            \item Solve $\hat{\bm{\beta}}_1^{(IPW-WEE)}$ from Equation (\ref{beta,IPW-WEE}), which contains the outcome model via $Y \sim A+\bm{X}$.
            \item Estimate $\hat{\tau}_1^{(IPW-WEE)}$ from Equation (\ref{IPW-WEE}).
        \end{enumerate}
    \item Repeat the ``two-step procedure" to estimate $\tau_0^{(IPW-WEE)}$ and finally obtain $\hat{\tau}^{(IPW-WEE)}$ as the main goal.
\end{enumerate}

Although IPW-DR is more robust than IPW-IPW, both IPW-IPW and IPW-DR estimators involve a joint effect of inverse weights on $\hat{\tau}_1$ from missingness and PS models. When the outcome model is wrong, and either estimated PS values are close to zero or missingness probabilities are close to 1, the extreme weights can directly enlarge the residual term $\Delta(Y_i,\bm{X}_i; \hat{\bm{\beta}}_1)=Y_i-E(Y_i|A_i=1,\bm{X}_i;\hat{\bm{\beta}}_1)$, which leads to the large bias and variance of estimated causal effects in finite samples.

In contrast, the major advantage of the IPW-WEE estimator is to reduce the joint effect of extreme weights. After applying WEE techniques, we find that even though $\bm{\beta}_1^{(IPW-WEE)}$ involves both inverse weights of the missingness and PS values, $\hat{\tau}_1^{(IPW-WEE)}$ only involves inverse weights of the missingness without the weights from PS model. In other words, IPW-WEE will not be directly affected by the extreme PS values by solving out $S(\bm{\beta}_1^{(IPW-WEE)})$. Another benefit of IPW-WEE is that it can be directly applied using weighted regression models based on Equation (\ref{beta,IPW-WEE}), by specifying ``weights" argument of the ``svyglm" function in R \cite{lumley2020package}. In summary, IPW-WEE keeps the same robust properties as IPW-DR, but becomes more resistant to the extreme weights in finite samples.


\section{TR Estimators}
\label{sec_TR}

In the previous section, we discussed the restriction of IPW estimators, which requires the missingness model to be correct. In this section, we develop TR robust estimators to protect against the misspecification of the missingness model. One is called ``TR-AIPW", extended from  \citet{williamson2012doubly,zhang} and another one is a newly developed TR-WEE method to reduce the effect of the extreme weights in finite samples. 

\subsection{Two Robust Estimators for Model Parameters}

\label{Sec_alpha_EE}

To improve the robust properties for TR estimators, we need to obtain more robust estimators for $(\bm{\alpha,\beta})$. One of ways to protect against the misspecification of the missingness model is to add an imputation model into the estimators of  $(\bm{\alpha,\beta})$.  We denote the estimator as $\hat{\bm{\alpha}}^{(EE)}$ which solves,
\begin{equation}
\begin{aligned}
\label{alpha_DR}
S({\bm{\alpha}}^{(EE)})=\sum_{i=1}^n \left\{\frac{1-R_i}{1-P_{R_i}(\bm{\hat{\gamma}})} U_i(A_i,\bm{X}_i;\bm{\alpha})-\frac{P_{R_i}(\bm{\hat{\gamma}})-R_i}{1-P_{R_i}(\bm{\hat{\gamma}})} \hat{m}_A (\bm{X}_i,Y_i) \right\}=0,
\end{aligned}    
\end{equation}
where $U_i(A_i,\bm{X}_i; \bm{\alpha}) =\bm{X}_i[A_i-E(A_i|\bm{X}_i;\bm{\alpha})]$ is  $(p+1) \times 1$ vector of the original score function from PS model for individual $i$, and $\hat{m}_A (\bm{X}_i,Y_i)=E[U_i(A_i,\bm{X}_i;\bm{\alpha})|\bm{X,Y};\bm{\hat{\delta}}]$ is the conditional expectation taken w.r.t $(\bm{A}|\bm{X,Y})$. Here, $(\hat{\bm{\gamma}}, \hat{\bm{\delta}})$ can be estimated by the missingness and imputation models, respectively, defined in Section \ref{sec_notation}. After we fit the imputation model, we can substitute $\hat{\bm{\delta}}$ for $\bm{\delta}$, so $\hat{m}_A (\bm{X}_i,Y_i)$ is the fitted value for the conditional expectation. If PS model is correct and either the missingness or the imputation model is correct, we prove $\bm{\hat{\alpha}}^{(EE)} \xrightarrow[]{\text{p}} \bm{\alpha^*}$ and its asymptotic normality holds as $n \to \infty$. The details of the proof are provided in S.1 of the supplementary material. 

Similarly,  we apply inverse weights of the missingness into the outcome model to adjust for the missingness, and the estimator of $\bm{\beta}$ is denoted as $\hat{\bm{\beta}}^{(EE)}$, which solves,
\begin{equation}
\begin{aligned}
\label{beta_DR_Miss}
S({\bm{\beta}}^{(EE)})=\sum_{i=1}^n \left\{\frac{1-R_i}{1-P_{R_i}(\bm{\hat{\gamma}})} V_{i}(Y_i,A_i, \bm{X}_i; \bm{\beta}) -\frac{P_{R_i}(\bm{\hat{\gamma}})-R_i}{1-P_{R_i}(\bm{\hat{\gamma}})} \hat{m}_Y (\bm{X}_i,Y_i) \right\}=0,
\end{aligned}    
\end{equation}
where $V_{i}(Y_i,A_i, \bm{X}_i; \bm{\beta}) =\bm{X}_i[Y_i-E(Y_i|A_i,\bm{X}_i;\bm{\beta})]$ is the $(p+2) \times 1$ vector of the original score function from the binary outcome model with the covariates and the exposure. Here, $\hat{m}_Y (\bm{X}_i,Y_i)=E[V_{i}(Y_i,A_i, \bm{X}_i; \bm{\beta})|\bm{X,Y;\delta}]$ is also taken w.r.t $(A|\bm{X},Y)$. After we fit the imputation model, we substitute $\hat{\bm{\delta}}$ for $\bm{\delta}$. If the outcome model is correctly specified and either the missingness or the imputation model is also correct, we prove $\bm{\hat{\beta}}^{(EE)} \xrightarrow[]{\text{p}} \bm{\beta^*}$ and its asymptotic normality holds as $n \to \infty$.  Notice that both $(\hat{\bm{\alpha}}^{(EE)},\hat{\bm{\beta}}^{(EE)})$  are just plugged-in values for IPW or TR estimators, instead of our main interest.

\subsection{TR-AIPW Estimator}
\label{Sec_TR_AIPW}

The ``TR-AIPW" estimator is constructed as a combination of two DR estimators to adjust for both missingness and confounding issues. The key differences between proposed TR-AIPW and previous DR estimators \cite{williamson2012doubly,zhang}) are that we improve the robust properties using a Bayesian approach in Equation (\ref{bayes}) to solve the issue when both missingness and imputation models are wrong.  We denote TR estimator as $\hat{\tau}_1^{(TR-AIPW)}$, written as:
\begin{equation}
\small
\begin{aligned}
\label{TR-AIPW}
\hat{\tau}_1^{(TR-AIPW)}
&=\frac{1}{n} \sum_{i=1}^n \frac{1-R_i}{1-P_{R_i}({\bm{\hat{\gamma}}})} \bm{Q_{i1}}(Y_i,A_i,\bm{X_i;\hat{\alpha}^{(EE)},\hat{\beta}^{(EE)}_1}) \\
& \indent -\frac{1}{n} \sum_{i=1}^n \frac{P_{R_i}({\bm{\hat{\gamma}}})-R_i}{1-P_{R_i}({\bm{\hat{\gamma}}})}E\left[ \bm{Q_{i1}}(Y_i,A_i,\bm{X_i;\hat{\alpha}^{(EE)},\hat{\beta}^{(EE)}_1}) \Bigg|\bm{X,Y;\hat{\delta}}\right] ,
\end{aligned}
\normalsize
\end{equation}
where  
$\bm{Q_{i1}}(Y_i,A_i,\bm{X_i;\hat{\alpha}^{(EE)},\hat{\beta}^{(EE)}})= \frac{A_i}{P_{A_i}(\hat{\bm{\alpha}}^{(EE)})}Y_i-\frac{A_i-P_{A_i}(\hat{\bm{\alpha}}^{(EE)})}{P_{A_i}(\hat{\bm{\alpha}}^{(EE)})}E(Y_i|A_i=1,\bm{X}_i;\hat{\bm{\beta}}^{(EE)}_1)$, and the subscript ``1" refers to the exposure group. Here, $P_{A_i}(\hat{\bm{\alpha}})$ is fitted PS values  and $E(Y_i|A_i=1,\bm{X}_i;\hat{\bm{\beta}}_1)$ is predicted responses for each individual.  $(\bm{\hat{\gamma}},\bm{\hat{\delta}})$ are estimated via the missingness and the imputation models, defined in Section \ref{sec_notation}, respectively. Here,  $E[\bm{Q_{i1}}(Y_i,A_i,\bm{X_i;\hat{\alpha}^{(EE)},\hat{\beta}^{(EE)}_1})|\bm{X,Y},\hat{\bm{\delta}}]$ is taken w.r.t $A|\bm{X},Y$ after plugging in $\bm{\hat{\delta}}$. $\hat{\bm{\alpha}}^{(EE)}$ are estimated from Equation (\ref{alpha_DR}) and $\hat{\bm{\beta}}_1^{(EE)}$ are estimated values on the exposure group from Equation (\ref{beta_DR_Miss}).


If the missingness model is correct, the expectation of the last term $\frac{P_{R_i}({\bm{\hat{\gamma}}})-R_i}{1-P_{R_i}({\bm{\hat{\gamma}}})}E\left[ \bm{Q_{i1}}(Y_i,A_i,\bm{X_i;\hat{\alpha}^{(EE)},\hat{\beta}^{(EE)}_1}) \Bigg|\bm{X,Y;\hat{\delta}}\right]$ is zero, due to MAR assumption. Notice that $\frac{1}{n} \sum_{i=1}^n\bm{Q_{i1}}(Y_i,A_i,\bm{X_i;\hat{\alpha},\hat{\beta}_1})$ is a DR estimator, which only requires either PS or the outcome model to be correct \cite{robins1994estimation,Bang,funk2011doubly}, so we know $E[\hat{\tau}_1^{(TR-AIPW)}]=E[Y^1]$. 

If the missingness model is wrong, but the imputation model is correct and either PS or the outcome model is correct,  $\{E[ \bm{Q_{i1}}(Y_i,A_i,\bm{X_i;\hat{\alpha}^{(EE)},\hat{\beta}^{(EE)}_1})|\bm{X,Y}]-E[ \bm{Q_{i1}}(Y_i,A_i,\bm{X_i;\hat{\alpha}^{(EE)},\hat{\beta}^{(EE)}_1})|\bm{X,Y;\hat{\delta}}]\}+E[Y^1]=0+E[Y^1]$. 

In the third case, if both missingness and imputation models are wrong, but PS and outcome models are correct, a Bayesian approach can be used to estimate $P(A=1|\bm{X},Y)$ in the imputation model. In this case, we can directly transfer the imputation model into functions of PS and the outcome models as:
\begin{equation}
\begin{aligned}
\label{bayes}
P(A=1|\bm{X},Y=y) =\frac{P(Y=y|A=1,\bm{X}) P(A=1|\bm{X})}{P(Y=y|\bm{X})} 
\end{aligned}    
\end{equation}
where $P(Y=y|\bm{X})$ can be simplified into two parts. When $y=1$, we know $P(Y=1|\bm{X})=E[E(Y|A,\bm{X})|\bm{X}]=P(Y=1|A=1,\bm{X})P(A=1|\bm{X})+P(Y=1|A=0,\bm{X})P(A=0|\bm{X})$. When $y=0$, we know $P(Y=0|\bm{X})=1-P(Y=1|\bm{X})$.  When both PS and outcome models are correct, applying the Bayesian approach helps us consistently estimate $P(A=1|\bm{X},Y=y)$, i.e. the imputation model w.r.t $\bm{\delta}$. Therefore, we can transform this third case back to the previously solved problem, same as the imputation model is correct and both PS and the outcome models are also correct.



In summary, as long as ``two out of three groups of models to be correct" (defined in Section \ref{sec_notation}), called ``TR group properties", TR-AIPW estimator for $\tau$ is consistent and its asymptotic normality holds as $n \to \infty$. The detailed proofs for the consistency and asymptotic properties of ``TR-AIPW" are shown in S.3 and S.4 of the supplementary material, respectively.




\subsection{TR-WEE Estimator}
\label{Sec_TR_WEE}

Although we provide $\hat{\tau}^{(TR-AIPW)}$ with a more robust property, TR-AIPW may still be affected by the joint effect of extreme weights in finite samples. In this subsection, we propose another TR estimator based on WEE, which is expected to be more stable after diminishing the joint effect of some extreme weights.  Similar to the idea of IPW-WEE, we construct TR-WEE in two steps. In the first step, we denote WEE estimator for the coefficients of the outcome model as $\hat{\bm{\beta}}_1^{(TR-WEE)}$ to solve the following equation: 
\begin{equation}
\begin{aligned}
\small
\label{beta,TR-WEE}
&S(\bm{\beta}_1^{(TR-WEE)}) \\
&= \sum_{i=1}^{n} \frac{1-R_i}{1-P_{R_i} (\bm{\hat{\gamma}})}\frac{\bm{X}_i}{P_{A_i}(\bm{\hat{\alpha}^{(EE)}})}  \left\{ A_i\Delta_i(Y_i,\bm{X}_i;\bm{\beta}_1)- E\left[A_i \Delta_i(Y_i,\bm{X}_i;\hat{\bm{\beta}}_1^{(EE)})\Bigg|\bm{X,Y;\hat{\delta}} \right] \right\} \\
&+ \sum_{i=1}^n E\left[ \bm{Q_{i1}}(Y_i,A_i,\bm{X_i;\hat{\alpha}^{(EE)},\hat{\beta}^{(EE)}_1}) \Bigg|\bm{X,Y;\hat{\delta}}\right]
=0,
\end{aligned}
\normalsize
\end{equation}
where $\Delta(Y_i,\bm{X}_i; \hat{\bm{\beta}}_1)=Y_i-E(Y_i|A_i=1,\bm{X}_i;\bm{\beta}_1)$ is the residual term and $E(Y_i|A_i=1,\bm{X}_i;\bm{\beta}_1)=\text{expit}(\bm{X}_i^T \bm{\beta}_1)$ is the outcome model for the exposure group with $\bm{X}_i$  including one intercept and $p$ covariates. Here, $(\bm{\hat{\alpha}^{(EE)}},\bm{\hat{\beta}^{(EE)}})$ are still estimated values from Equations (\ref{alpha_DR})-(\ref{beta_DR_Miss}) and $\bm{Q_{i1}}(Y_i,A_i,\bm{X_i;\hat{\alpha}^{(EE)},\hat{\beta}^{(EE)}})= \frac{A_i}{P_{A_i}(\hat{\bm{\alpha}}^{(EE)})}Y_i-\frac{A_i-P_{A_i}(\hat{\bm{\alpha}}^{(EE)})}{P_{A_i}(\hat{\bm{\alpha}})^{(EE)}}E(Y_i|A_i=1,\bm{X}_i;\hat{\bm{\beta}}^{(EE)}_1)$ is DR estimator for individual $i=1,2,\dots,n$.

If the outcome model is correct and either the missingness or the imputation model is correct, we prove that $\hat{\bm{\beta}}_1^{(TR-WEE)} \xrightarrow[]{\text{p}} \bm{\beta}_1^*$ and its asymptotic normality holds as $n \to \infty$. Alternatively, if both missingness and imputation models are wrong, but PS and outcome models are correct, we can apply the Bayesian approach in Equation (\ref{bayes}) to consistently estimate $P(A=1|\bm{X},Y)$. 

In the second step, we denote $\hat{\tau}_1^{(TR-WEE)}$ as TR-WEE estimator for $\tau_1$, written as:
\begin{equation}
\begin{aligned}
\label{TR-WEE}
\hat{\tau}_1^{(TR-WEE)}=
&\frac{1}{n} \sum_{i=1}^n \frac{1-R_i}{1-P_{R_i}({\bm{\hat{\gamma}}})} \left[E(Y_i|A_i=1,\bm{X}_i; \hat{\bm{\beta}}_1^{(TR-WEE)})-E(Y_i|A_i=1,\bm{X}_i;\hat{\bm{\beta}}^{(EE)}_1) \right] \\
&+\frac{1}{n} \sum_{i=1}^n E(Y_i|A_i=1,\bm{X}_i; \hat{\bm{\beta}}_1^{(EE)}),
\end{aligned}
\end{equation}
where  $E \left[Y_i|A_i=1,\bm{X}_i,\hat{\bm{\beta}}_1^{(TR-WEE)} \right]=\text{expit}(\bm{X}_i^T \hat{\bm{\beta}}_1^{(TR-WEE)})$ is the predicted response for each individual after plugging in $\hat{\bm{\beta}}_1^{(TR-WEE)}$. 


In summary, as long as ``two out of three groups of models to be correct",   $\hat{\tau}^{(TR-WEE)}= \frac{\hat{\tau}_1^{(TR-WEE)}/(1-\hat{\tau}_1^{(TR-WEE)})}{\hat{\tau}_0^{(TR-WEE)}/(1-\hat{\tau}_0^{(TR-WEE)})} \xrightarrow[]{\text{p}} \tau$ and its asymptotic normality holds as $n \to \infty$. The detailed proof for the connection between two TR estimators is provided in S.3.2 and the asymptotic normality for $\hat{\tau}$ is also shown in S.4 of the supplementary material.

We revise the previous algorithm for IPW-WEE and provide a new algorithm to conduct the proposed TR-WEE estimator in five stages:
\begin{enumerate}
    \item Fit the missingness model and obtain $\bm{\hat{\gamma}}$ via $R\sim \bm{X}+Y$ as the plugged-in values in Step 4 (a);
    \item Fit the imputation model and obtain $\hat{\delta}$ via $A \sim \bm{X}+Y$ as the plugged-in values in Step 4 (a);
    \item Solve $\hat{\bm{\alpha}}^{(EE)}, \hat{\bm{\beta}}^{(EE)}$ from Equation (\ref{alpha_DR}) and Equation (\ref{beta_DR_Miss}), respectively, as the plugged-in values. 
    \item Conduct the ``two-step procedure" to avoid joint effects of the extreme weights. 
        \begin{enumerate}
            \item Solve $\hat{\bm{\beta}}_1^{(TR-WEE)}$ from Equation (\ref{beta,TR-WEE}), which contains the outcome model via $Y \sim A+X$.
            \item Estimate $\hat{\tau}_1^{(TR-WEE)}$ from Equation (\ref{TR-WEE}).
        \end{enumerate}
    \item Repeat the ``two-step procedure" to estimate $\tau_0^{(TR-WEE)}$ and finally obtain $\hat{\tau}^{(TR-WEE)}$ as the main goal.
\end{enumerate}

Comparing TR with IPW estimators, two TR estimators allow more flexible conditions to achieve consistency and asymptotic normality, which further protect against the misspecification of the missingness or the imputation model. The key reasons are that the estimating equations are revised and more robust plugged-in values are used, i.e. $(\hat{\bm{\alpha}}^{(EE)}, \hat{\bm{\beta}}^{(EE)})$. The comparison among different estimators is summarized in Table \ref{table: DR_TR}.
\begin{table}[ht]
    \centering
    \caption{Comparison of different IPW, DR, and TR estimators}
    \label{table: DR_TR}
    \begin{tabular}{cccc}
    \hline
    Method & Used Models & Type of Analysis & Requirement of Consistency \\
    \hline
    $\hat{{\alpha}}^{(EE)}$ & MS, Imp, PS & Associational  & (MS/Imp  + PS ); Bayes: (PS + OR) \\
    $\hat{{\beta}}^{(EE)}$ & MS, Imp, OR & Associational &  (MS/Imp  + OR); Bayes: (PS + OR)  \\
    $\hat{{\tau}}^{(IPW-IPW)}$ & MS, PS & Causal & MS + PS . No Imp model. \\
    $\hat{{\tau}}^{(IPW-WEE)}$ and $\hat{{\tau}}^{(IPW-DR)}$ & MS, PS, OR & Causal & MS  + PS/OR . No Imp model. \\
    $\hat{{\tau}}^{(TR-WEE)}$ and  $\hat{{\tau}}^{(TR-AIPW)}$ & MS, Imp, PS, OR & Causal & (MS/Imp  + PS/OR ); Bayes: (PS + OR) \\
    \hline
    \end{tabular}
    \footnotetext{
    MS, Imp, PS, and OR, refer to the missingness, imputation, PS, and outcome models, respectively. ``$\slash$" means one of the models needs to be correct for consistency. ``$+$" means at least two models should be correct. Bayes: alternative Bayesian approach when MS and Imp models are wrong, but PS and OR models are correct.
   }
\end{table}

To compare with TR-WEE with TR-AIPW, even though TR-AIPW involves both inverse weights of the missingness and PS values in Equation (\ref{TR-AIPW}),  TR-WEE estimator only involves weights of the missingness without any weights from PS model in Equation (\ref{TR-WEE}). In summary, TR-WEE not only keeps the same robust properties as TR-AIPW, but also reduces the joint effect of extreme weights on the bias and variance compared with TR-AIPW

\section{Simulation Study}
\label{sec_simulation}

\subsection{Model Setting}

A simulation study is conducted to investigate the finite sample performance of the proposed methods. The number of simulation replications is $N=500$, the number of bootstraps for obtaining the confidence intervals is $B=2000$, and the sample size is $n=1000$ in each data generation. Three covariates $X_1,X_2,X_3$ are generated from $N(0,1)$ independently. The binary exposure $A$, outcome $Y$, and missing indicator $R$ are generated by the following three models:
\begin{itemize}
\item For PS model: $\text{logit}\{P(A=1|\bm{X})\}=-0.2+0.9X_1+X_2+0.8X_3$,
\item For the outcome model: $\text{logit}\{P(Y=1|A, \bm{X})\}=0.7+A+0.5X_1+0.9X_2+0.7X_3$,
\item For the missingness model: $\text{logit}\{P(R=1|\bm{X},Y)\}=-0.5+0.6X_1+0.7X_2+0.8X_3+0.5Y,$
\end{itemize}
The missing rate is about 47.5\% and due to the non-linear link,  the true causal effect can be found numerically as $\tau=2.201$. The numerical method to obtain the true causal effect is provided in S.5 of the supplementary material. Single imputation (SI), multiple imputations chained equations (MICE), and the proposed IPW and TR estimators are compared to estimate the causal odds ratio when the exposure is MAR. In MICE, $M=10$ imputation times are chosen for imputing the missing data. Then, DR estimators for $\tau$ can be constructed on each imputed dataset, called ``DR-MICE".

\subsection{Estimation for IPW Method}
\label{Simu_DR_TR}

To evaluate the performance of different estimators, the description of measurements is listed below: (1) $\text{Bias}=\frac{1}{N} \sum_{j=1}^N \hat{{\tau}_j}-\tau$, where $\hat{{\tau}}_j$ is the point estimate in the $j^{th}$ replication, $j=1,2,\dots,N$; (2) The bias rate is of main interest, defined as $\text{bias rate}=\text{bias}/\tau$; (3) The empirical standard error (ESE): $\sqrt{\frac{1}{N-1} \sum_{j=1}^N(\hat{{\tau}_j}-\Bar{{\tau}})^2}$, which is also the sampling standard error; (4) The RMSE is the square root of mean squared error, written as $\sqrt{\frac{1}{N-1}\sum_{j=1}^N(\hat{{\tau}}_j-\tau)^2}$. (4) Median BSE: the median values of bootstrap standard errors (5)  CI Percentile: the coverage rates of 95\% CI bootstrap percentiles.

Since we do not need to specify the imputation model for IPW strategy, we first compare IPW methods with eight different model specifications of the missingness, PS, and outcome models. We will intentionally specify the wrong model by disregarding $X_3$ if  PS, or outcome model is incorrectly specified. If the missingness or the imputation model is wrong, researchers may ignore the effect of the outcome, so we will disregard $Y$ from the model following the setup of the simulation studies from \citet{zhang}.

\begin{figure}[ht]
\centering
\includegraphics[width=0.9\textwidth,height=9cm]{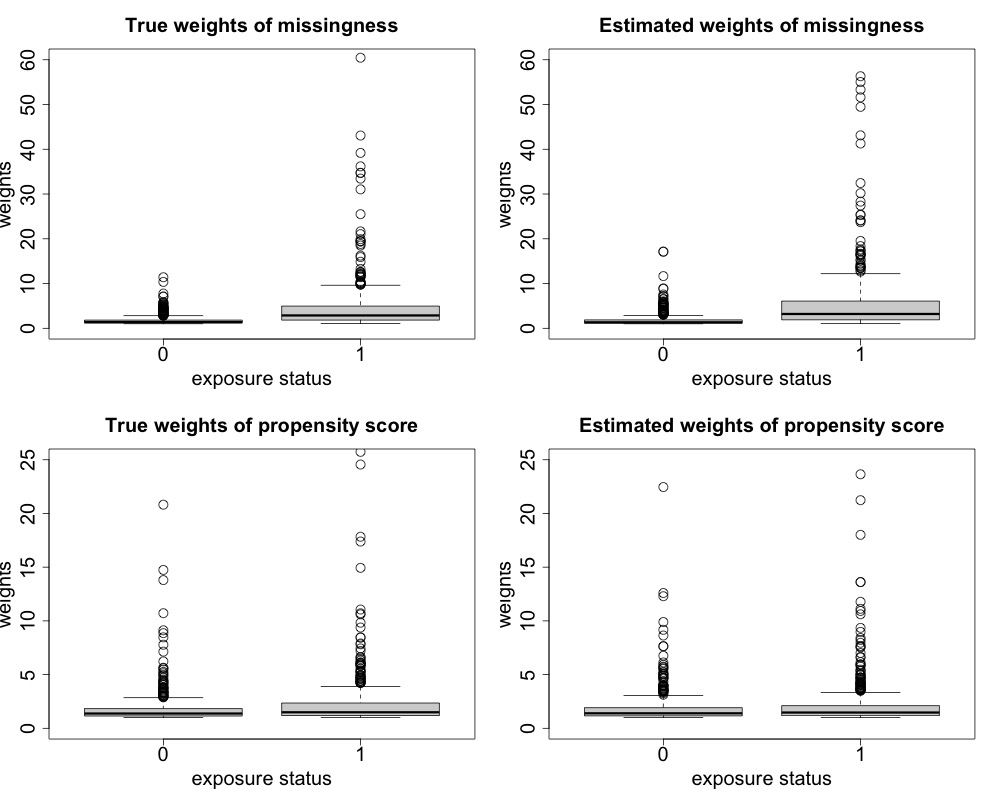}
\caption{Boxplot of inverse weights in one simulation study when all models are correct. The weights of the missingness and PS are provided on the first and the second rows, respectively. 
}
\label{fig:box_weight} 
\end{figure}

Figure \ref{fig:box_weight} is the boxplot of the inverse weights of the missingness and PS values when all models are correct, which shows most estimated weights are close to the true weights. In the simulation studies, we find that some weights from the missingness model are larger than 60 due to the high missing rates. Meanwhile, some true weights of PS values are also larger than 25, which may also largely affect the estimation of the causal effect. That motivates us to consider IPW-WEE or TR-WEE approach to reduce the joint effect of extreme weights from both missingness and PS models.

The main results for IPW strategies are presented in Table \ref{table:ipw20}, which shows IPW-WEE method is more efficient than IPW-IPW and IPW-DR estimators in finite samples. The boxplot of point estimation is displayed in Figure \ref{fig:boxplot_ipw_sim} of the Appendix to describe the distribution of $\hat{\tau}$. Figure \ref{fig:boxplot_ipw_sim} clearly shows that both IPW-IPW and IPW-DR generate more outliers and larger variance of estimated results than IPW-WEE.  To be more specific,  from Table \ref{table:ipw20},  we find that IPW-WEE generates the smallest ESE and RMSE in different cases among other estimators. As we have mentioned before, the joint effect of the missingness and PS models can largely increase the variation from Figure \ref{fig:box_weight}, but IPW-WEE is resistant to those extreme weights, which shows a more stable results in terms of bias and ESE than others. 

For robustness, both IPW-DR and IPW-WEE can protect against the misspecification of either PS or OR model. For example, when the PS model is wrong, but both the missingness and outcome models are correct, IPW-IPW creates a huge bias, compared with both IPW-DR and IPW-WEE with relatively small bias. In that case, IPW-IPW cannot properly estimate the true causal effect because IPW-IPW estimator requires both missingness and PS models to be correct. Even though all models are correct, both IPW-DR and IPW-WEE lead to smaller bias rates than IPW-IPW because IPW-IPW estimator can be easily affected by the extreme weights. 

In terms of standard errors, we provide median BSE because the extreme values can largely affect the mean values of BSE. From Table \ref{table:ipw20}, we find that the median BSE is closer to ESE using IPW-WEE method. In comparison, IPW-IPW and IPW-DR can generate a larger gap between ESE and median BSE because their estimated values are largely affected by extreme weights. When all models are correct, all IPW estimators lead to the coverage rates of CI close to 95\%. If either the missingness or PS model is wrong, IPW-WEE can still generate coverage rates close to 95\%, but IPW-IPW generates much smaller coverage rates, and IPW-DR slightly overestimates the coverage rates.  When both PS and the outcome models are wrong or even all models are wrong, all estimators will yield huge bias and very low coverage rates because the requirement of the model specification is not satisfied. 

In summary, our proposed IPW-WEE estimator results in smaller RMSE, and when some extreme weights occur, IPW-WEE method can help us reduce the joint effect of those extreme weights in finite samples.

\begin{table}[ht]
\centering
\caption{Performance of IPW estimators in different model specification} 
\label{table:ipw20}
\begin{tabular}{lcccccccc}
 \hline
 & Bias & Bias Rate & ESE & Median BSE & RMSE & CI Percentile \\ 
  \hline
All True\cmark &  &  &  &  &  &  \\ 
  IPW-IPW & 0.264 & 12.007 & 12.311 & 1.843 & 12.314 & 95.0 \\ 
  IPW-DR & 0.112 & 5.076 & 1.388 & 0.673 & 1.392 & 96.0 \\ 
  IPW-WEE & 0.086 & 3.894 & 0.600 & 0.601 & 0.607 & 95.6 \\ 
    \hdashline
  MS PS\cmark OR \xmark &  &  &  &  &  &  \\ 
  IPW-IPW & 0.264 & 12.007 & 12.311 & 1.843 & 12.314 & 95.0 \\ 
  IPW-DR & 0.160 & 7.270 & 1.666 & 0.703 & 1.674 & 97.0 \\ 
  IPW-WEE & 0.132 & 5.990 & 0.635 & 0.662 & 0.648 & 96.6 \\ 
    \hdashline
  MS OR\cmark PS \xmark &  &  &  &  &  &  \\ 
  IPW-IPW & 1.904 & 86.483 & 4.003 & 2.127 & 4.433 & 67.2 \\ 
  IPW-DR & 0.061 & 2.763 & 0.836 & 0.598 & 0.838 & 96.0 \\ 
  IPW-WEE & 0.071 & 3.243 & 0.558 & 0.583 & 0.563 & 95.8 \\ 
    \hdashline
  PS OR\cmark MS \xmark &  &  &  &  &  &  \\ 
  IPW-IPW & 0.815 & 37.007 & 4.245 & 1.939 & 4.323 & 93.6 \\ 
  IPW-DR & 0.167 & 7.597 & 1.314 & 0.746 & 1.324 & 97.0 \\ 
  IPW-WEE & 0.157 & 7.155 & 0.658 & 0.650 & 0.676 & 96.2 \\ 
    \hdashline
  MS\cmark PS OR \xmark &  &  &  &  &  &  \\ 
  IPW-IPW & 1.904 & 86.483 & 4.003 & 2.127 & 4.433 & 67.2 \\ 
  IPW-DR & 1.391 & 63.218 & 1.204 & 1.169 & 1.840 & 60.0 \\ 
  IPW-WEE & 1.393 & 63.301 & 1.017 & 1.075 & 1.725 & 57.0 \\ 
    \hdashline
  PS\cmark MS OR \xmark &  &  &  &  &  &  \\ 
  IPW-IPW & 0.815 & 37.007 & 4.245 & 1.939 & 4.323 & 93.6 \\ 
  IPW-DR & 0.246 & 11.183 & 2.033 & 0.750 & 2.048 & 97.2 \\ 
  IPW-WEE & 0.204 & 9.264 & 0.695 & 0.702 & 0.724 & 96.2 \\ 
    \hdashline
  OR\cmark MS PS \xmark &  &  &  &  &  &  \\ 
  IPW-IPW & 1.833 & 83.255 & 2.718 & 2.143 & 3.278 & 68.8 \\ 
  IPW-DR & 0.137 & 6.217 & 0.844 & 0.654 & 0.855 & 95.8 \\ 
  IPW-WEE & 0.142 & 6.467 & 0.608 & 0.628 & 0.625 & 96.0 \\ 
    \hdashline
  All False \xmark &  &  &  &  &  &  \\ 
  IPW-IPW & 1.833 & 83.255 & 2.718 & 2.143 & 3.278 & 68.8 \\ 
  IPW-DR & 1.378 & 62.599 & 1.245 & 1.219 & 1.857 & 65.2 \\ 
  IPW-WEE & 1.384 & 62.898 & 1.072 & 1.113 & 1.751 & 61.4 \\ 
   \hline
\end{tabular}
\footnotetext{ 
      The first column lists which model is correct (\cmark) or wrong (\xmark). MS, PS, and OR, refer to the missingness, PS, and the outcome model, respectively. 
    }
\end{table}

\subsection{Estimation for DR or TR Method}

In the next simulation study, we compare the performance of DR with TR estimators. Similar to the setup of IPW estimator, if PS or outcome model is wrong, we will disregard $X_3$ from the model.   If the missingness or imputation model is wrongly specified, we will remove $Y$ from that model.  The main result with $n=1000$ are shown as Tables \ref{table:beta_tr20_p1} and \ref{table:beta_tr20_p2} in two parts with the sixteen different model specifications of the imputation, the missingness, PS, and the outcome models. we also provide Figure \ref{fig:boxplot_tr_sim} in Appendix as the boxplot of point estimation using DR and TR methods.

\begin{table}[ht]
\centering
\caption{Performance of DR and TR estimators in different model specifications [part 1]} 
\label{table:beta_tr20_p1}
\begin{tabular}{lcccccc}
\hline
 & Bias & Bias Rate & ESE & Median BSE & RMSE & CI Percentile \\ 
  \hline
All True\cmark &  &  &  &  &  &  \\ 
  DR-SI & 0.085 & 3.846 & 0.706 & 0.633 & 0.711 & 98.4 \\ 
  DR-MICE & 0.103 & 4.692 & 1.762 & 0.577 & 1.765 & 97.2 \\ 
  TR-AIPW & 0.153 & 6.956 & 1.176 & 0.657 & 1.186 & 96.8 \\ 
  TR-WEE & 0.084 & 3.836 & 0.588 & 0.614 & 0.594 & 96.2 \\ 
  \hdashline
  MS PS OR\cmark Imp \xmark &  &  &  &  &  &  \\ 
  DR-SI & -0.491 & -22.300 & 0.432 & 0.329 & 0.654 & 69.8 \\ 
  DR-MICE & -0.495 & -22.503 & 0.469 & 0.289 & 0.682 & 64.6 \\ 
  TR-AIPW & 0.162 & 7.376 & 1.066 & 0.660 & 1.078 & 97.8 \\ 
  TR-WEE & 0.108 & 4.892 & 0.595 & 0.643 & 0.605 & 95.8 \\ 
    \hdashline
  MS PS Imp\cmark OR \xmark &  &  &  &  &  &  \\ 
  DR-SI & 0.089 & 4.046 & 0.749 & 0.673 & 0.754 & 98.8 \\ 
  DR-MICE & 0.070 & 3.180 & 0.704 & 0.618 & 0.707 & 97.6 \\ 
  TR-AIPW & 0.093 & 4.242 & 0.790 & 0.678 & 0.796 & 96.6 \\ 
  TR-WEE & 0.129 & 5.855 & 0.616 & 0.676 & 0.630 & 95.6 \\ 
    \hdashline
  MS OR Imp\cmark PS \xmark &  &  &  &  &  &  \\ 
  DR-SI & 0.038 & 1.746 & 0.974 & 0.673 & 0.975 & 98.4 \\ 
  DR-MICE & 0.025 & 1.142 & 0.856 & 0.618 & 0.856 & 96.4 \\ 
  TR-AIPW & 0.050 & 2.255 & 0.698 & 0.678 & 0.700 & 96.0 \\ 
  TR-WEE & 0.066 & 2.996 & 0.541 & 0.676 & 0.545 & 96.6 \\ 
    \hdashline
  PS OR Imp\cmark MS \xmark &  &  &  &  &  &  \\ 
  DR-SI & 0.085 & 3.855 & 0.708 & 0.630 & 0.713 & 98.4 \\ 
  DR-MICE & 0.103 & 4.680 & 1.772 & 0.577 & 1.775 & 97.2 \\ 
  TR-AIPW & 0.161 & 7.294 & 1.320 & 0.675 & 1.330 & 96.6 \\ 
  TR-WEE & 0.079 & 3.611 & 0.616 & 0.633 & 0.621 & 95.7 \\ 
    \hdashline
  MS PS\cmark Imp OR \xmark &  &  &  &  &  &  \\ 
  DR-SI & -0.504 & -22.910 & 0.424 & 0.350 & 0.659 & 73.0 \\ 
  DR-MICE & -0.511 & -23.206 & 0.487 & 0.311 & 0.706 & 68.0 \\ 
  TR-AIPW & 0.087 & 3.961 & 0.778 & 0.662 & 0.783 & 97.2 \\ 
  TR-WEE & 0.117 & 5.331 & 0.603 & 0.651 & 0.614 & 96.8 \\ 
    \hdashline
  MS OR\cmark Imp PS \xmark &  &  &  &  &  &  \\ 
  DR-SI & -0.561 & -25.500 & 0.499 & 0.285 & 0.751 & 60.4 \\ 
  DR-MICE & -0.565 & -25.676 & 0.533 & 0.238 & 0.777 & 49.0 \\ 
  TR-AIPW & 0.044 & 2.018 & 0.666 & 0.562 & 0.668 & 96.6 \\ 
  TR-WEE & 0.060 & 2.736 & 0.533 & 0.560 & 0.537 & 97.2 \\ 
      \hdashline
  MS Imp\cmark PS OR \xmark &  &  &  &  &  &  \\ 
  DR-SI & 1.316 & 59.779 & 1.150 & 1.007 & 1.747 & 57.0 \\ 
  DR-MICE & 1.285 & 58.366 & 0.992 & 0.915 & 1.623 & 49.8 \\ 
  TR-AIPW & 1.337 & 60.750 & 1.009 & 0.971 & 1.675 & 53.6 \\ 
  TR-WEE & 1.361 & 61.846 & 0.896 & 0.947 & 1.630 & 51.0 \\ 
   \hline
\end{tabular}
 \footnotetext{
      The first column lists which model is correct (\cmark) or wrong (\xmark).  MS, Imp, PS, and OR, refer to the missingness, the imputation, PS, and the outcome model, respectively. 
    }
    
\end{table}

From Table \ref{table:beta_tr20_p1}, when all models are correct, we find that both SI and MICE lead to higher ESE and RMSE than TR-WEE estimators. Both imputation-based approaches also overestimate the coverage rates of CIs compared with the two TR estimators. If the imputation model is wrong, but the missingness, PS, and outcome models are correct, both DR-SI and DR-MICE generate larger than 20\% bias and low coverage rates of CI because of wrongly imputed data. Notice that imputation-based methods essentially require the imputation model to be correct, plus either PS or the outcome model to be correct, which is more restrictive than TR estimators. Compared with the two imputation methods, MICE performs better than SI because its coverage rates are closer to 95\%. 

To study the robustness of TR-WEE, Tables \ref{table:beta_tr20_p1}-\ref{table:beta_tr20_p2} show that TR estimators are consistent with true causal odds ratio if ``two out of three groups of models are correct". To be more specific, when one model is correctly specified from either the missingness or the imputation model, and another model is correctly specified from either PS or the outcome model, TR-WEE estimators can generate bias rates around 5\%. When both imputation and missingness models are wrong, but PS and outcome models are correct, we can apply the alternative Bayesian approach to correctly estimate the imputation model, so that TR-WEE estimator still can generate about 5\% bias. On the other hand, when both PS and outcome models are wrong or all models are wrong, the bias rates dramatically increase  using all estimators  because the requirement of the model specification is not satisfied.

In terms of the standard errors, when all models are correctly specified, the median BSE is closer to ESE using TR-WEE method than the imputation-based approach or TR-AIPW, which will result in a much larger gap between ESE and median BSE. In most scenarios, TR-WEE results in a smaller RMSE compared with other approaches.

To compare within two TR estimators, even though TR-AIPW contains the same robust properties as TR-WEE, TR-WEE shows more efficient results in different cases.  Figure \ref{fig:boxplot_tr_sim} shows that TR-WEE method contains a more narrow interquartile range, which results in a smaller variance than TR-AIPW. Additionally, the coverage rates of CI using  TR-WEE are closer to 95\% than CI using TR-AIPW when PS model is correct.

In summary, two TR estimators become more robust than the previous two-layer DR estimator and the imputation-based method. Compared with the properties of IPW estimators, TR method only requires ``two out of three groups of models to be correct", which can further protect against the misspecification of either the missingness or the imputation models. As TR group properties hold, the inference results are also valid with nearly 95\% coverage rates of CI. If we compare within the two types of TR estimators, TR-WEE estimator shows more efficient results and becomes more resistant to extreme weights than TR-AIPW in finite samples.   




\begin{table}[ht]
\centering
\caption{Performance of DR and TR estimators in different model specifications [part 2]} 
\label{table:beta_tr20_p2}
\begin{tabular}{lcccccc}
  \hline
 & Bias & Bias Rate & ESE & Median BSE & RMSE & CI Percentile \\ 
  \hline
PS OR\cmark MS Imp \xmark &  &  &  &  &  &  \\ 
  DR-SI & -0.491 & -22.315 & 0.433 & 0.329 & 0.655 & 69.7 \\ 
  DR-MICE & -0.495 & -22.505 & 0.470 & 0.290 & 0.683 & 64.5 \\ 
  TR-AIPW & 0.157 & 7.132 & 1.064 & 0.682 & 1.075 & 96.8 \\ 
  TR-WEE & 0.108 & 4.928 & 0.630 & 0.658 & 0.639 & 95.6 \\ 
    \hdashline
  PS Imp\cmark MS OR \xmark &  &  &  &  &  &  \\ 
  DR-SI & 0.088 & 4.013 & 0.750 & 0.671 & 0.755 & 98.8 \\ 
  DR-MICE & 0.069 & 3.138 & 0.704 & 0.618 & 0.708 & 97.4 \\ 
  TR-AIPW & 0.090 & 4.082 & 0.832 & 0.699 & 0.837 & 96.2 \\ 
  TR-WEE & 0.117 & 5.313 & 0.653 & 0.688 & 0.664 & 95.8 \\ 
    \hdashline
  OR Imp\cmark MS PS \xmark &  &  &  &  &  &  \\ 
  DR-SI & 0.038 & 1.746 & 0.974 & 0.605 & 0.975 & 98.4 \\ 
  DR-MICE & 0.025 & 1.142 & 0.856 & 0.543 & 0.856 & 96.4 \\ 
  TR-AIPW & 0.053 & 2.422 & 0.703 & 0.593 & 0.705 & 94.8 \\ 
  TR-WEE & 0.062 & 2.808 & 0.557 & 0.577 & 0.560 & 96.0 \\ 
    \hdashline
  MS\cmark Imp PS OR \xmark &  &  &  &  &  &  \\ 
  DR-SI & 0.383 & 17.412 & 0.642 & 0.521 & 0.747 & 93.6 \\ 
  DR-MICE & 0.257 & 11.692 & 3.137 & 0.444 & 3.147 & 88.0 \\ 
  TR-AIPW & 1.328 & 60.320 & 0.990 & 0.968 & 1.656 & 54.0 \\ 
  TR-WEE & 1.347 & 61.219 & 0.881 & 0.935 & 1.610 & 51.6 \\ 
    \hdashline
  PS\cmark MS Imp OR \xmark &  &  &  &  &  &  \\ 
  DR-SI & -0.504 & -22.910 & 0.424 & 0.350 & 0.659 & 73.2 \\ 
  DR-MICE & -0.511 & -23.206 & 0.487 & 0.311 & 0.706 & 68.2 \\ 
  TR-AIPW & 0.256 & 11.641 & 0.921 & 0.801 & 0.956 & 97.0 \\ 
  TR-WEE & 0.292 & 13.282 & 0.739 & 0.782 & 0.795 & 94.6 \\ 
    \hdashline
  OR\cmark MS Imp PS \xmark &  &  &  &  &  &  \\ 
  DR-SI & -0.561 & -25.500 & 0.499 & 0.283 & 0.751 & 60.6 \\ 
  DR-MICE & -0.565 & -25.676 & 0.533 & 0.237 & 0.777 & 48.4 \\ 
  TR-AIPW & 0.261 & 11.862 & 0.779 & 0.727 & 0.821 & 96.0 \\ 
  TR-WEE & 0.226 & 10.264 & 0.644 & 0.671 & 0.683 & 95.4 \\ 
    \hdashline
  Imp\cmark MS PS OR \xmark &  &  &  &  &  &  \\ 
  DR-SI & 1.316 & 59.779 & 1.150 & 1.007 & 1.747 & 57.2 \\ 
  DR-MICE & 1.285 & 58.366 & 0.992 & 0.915 & 1.623 & 49.8 \\ 
  TR-AIPW & 1.343 & 61.013 & 1.033 & 1.008 & 1.694 & 56.4 \\ 
  TR-WEE & 1.371 & 62.272 & 0.933 & 0.984 & 1.658 & 52.6 \\ 
    \hdashline
  All False \xmark &  &  &  &  &  &  \\ 
  DR-SI & 0.383 & 17.412 & 0.642 & 0.520 & 0.747 & 93.6 \\ 
  DR-MICE & 0.257 & 11.692 & 3.137 & 0.444 & 3.147 & 88.2 \\ 
  TR-AIPW & 1.641 & 74.574 & 1.189 & 1.244 & 2.027 & 51.0 \\ 
  TR-WEE & 1.664 & 75.576 & 1.099 & 1.195 & 1.994 & 47.4 \\ 
   \hline
\end{tabular}
 \footnotetext{
      The first column lists which model is correct (\cmark) or wrong (\xmark).  MS, Imp, PS, and OR, refer to the missingness, the imputation, PS, and the outcome model, respectively. 
    }
    
\end{table}

\section{Application}
\subsection{Background}

In this section, we conduct an application on a COVID-19 dataset to study the causal effect of cardiovascular disease (CVD) on the mortality of COVID-19 patients from Brazil, using different methods and comparing their performance. COVID-19 is a contagious disease that emerged in Wuhan, China in December 2019 and spread worldwide rapidly. As reported by  \citet{bansal}, it has been shown that the presence of CVD is associated with significantly worse mortality in COVID-19 patients. However, an analysis based on European data finds no significant association between CVD and higher mortality rates for COVID-19 patients \cite{di}. 

The data that we analyze is collected from September 20th to December 1st, 2020 during the second wave of the pandemic in Brazil. The total sample size is 927, and of these, 222 patients died from COVID-19. All patients in this dataset were exposed to COVID-19 disease, but 162 patients were missing information about CVD, so the missing rate is 17.5\%. Other demographic and clinical variables were collected including age, sex, and diabetes as major confounders with no missing data \cite{leon2015diabetes}.  The summary statistics are provided in Table \ref{table:summaryapply}. The tests of independence between mortality and covariates are based on $\chi^2$ test for categorical variables and the Mann-Whitney test for continuous variables, respectively. Based on Table \ref{table:summaryapply}, we only find the statistical association between mortality and age, which is a well-known risk factor.


\begin{table}[ht]
\centering
\caption{Summary table of COVID-19 data Stratified by mortality}
\label{table:summaryapply}
\begin{tabular}{lcccc}
   \hline
Covariate & Full Sample (n=927) & Alive (n=705) & Died (n=222) & p-value \\ 
   \hline
\textbf{CVD} &  &  &  & 0.25 \\ 
  ~~~No & 284 (37) & 224 (38) & 60 (33) &  \\ 
  ~~~Yes & 481 (63) & 361 (62) & 120 (67) &  \\ 
  ~~~Missing & \textbf{162} & \textbf{120} & \textbf{42} &  \\ 
  \hdashline
  \textbf{age} &  &  &  & \textbf{$<$0.001} \\ 
  ~~~Mean (sd) & 67 (17.1) & 64.7 (17.3) & 74.3 (14.3) &  \\ 
  ~~~Median (Min,Max) & 70 (0,107) & 67 (0,99) & 76 (20,107) &  \\ 
    \hdashline
  \textbf{sex} &  &  &  & 0.44 \\ 
  ~~~female & 464 (50) & 358 (51) & 106 (48) &  \\ 
  ~~~male & 463 (50) & 347 (49) & 116 (52) &  \\ 
    \hdashline
  \textbf{diabetes} &  &  &  & 0.18 \\ 
  ~~~No & 382 (41) & 282 (40) & 100 (45) &  \\ 
  ~~~Yes & 545 (59) & 423 (60) & 122 (55) &  \\ 
   \hline
\end{tabular}
 \footnotetext{
      CVD: cardiovascular disease. For the continuous variable, it shows the mean, se, min, max, and median. For the categorical variable, its frequency and percentage of total samples are provided in the bracket. 
    }
\end{table}

In such observational studies, we cannot manipulate the exposure status for patients, so confounding issues may exist. When the CVD status is missing, estimating the causal effect of CVD (as a binary exposure) on mortality (as a binary outcome) is challenging. That motivates us to apply proposed IPW and TR estimators to adjust for both missing and confounding issues. In addition, we also compare the performance of IPW and TR estimators with imputation-based methods in the application.  From the simulation study, since SI does not perform better than MICE, and IPW-IPW is easily affected by some extreme weights, we do not present them in the application study. 

Before we construct the estimators, we have to assume MAR and SITA assumptions based on the background knowledge in the application study. Typically, older COVID-19 patients with diabetes may be more likely to report whether they have CVD at the same time. Moreover, deceased patients are also more likely to be recorded on their health status before their death, than others who survive. That means assuming MCAR in this study is not reasonable because the missing mechanism may be affected by other covariates and the outcome, so we adopt MAR assumption in this case, i.e. assuming the missing indicator for CVD is independent of CVD itself given age, sex, diabetes, and mortality. 

\begin{figure}[ht]
\centering
\includegraphics[width=0.7\textwidth,height=2.4cm]{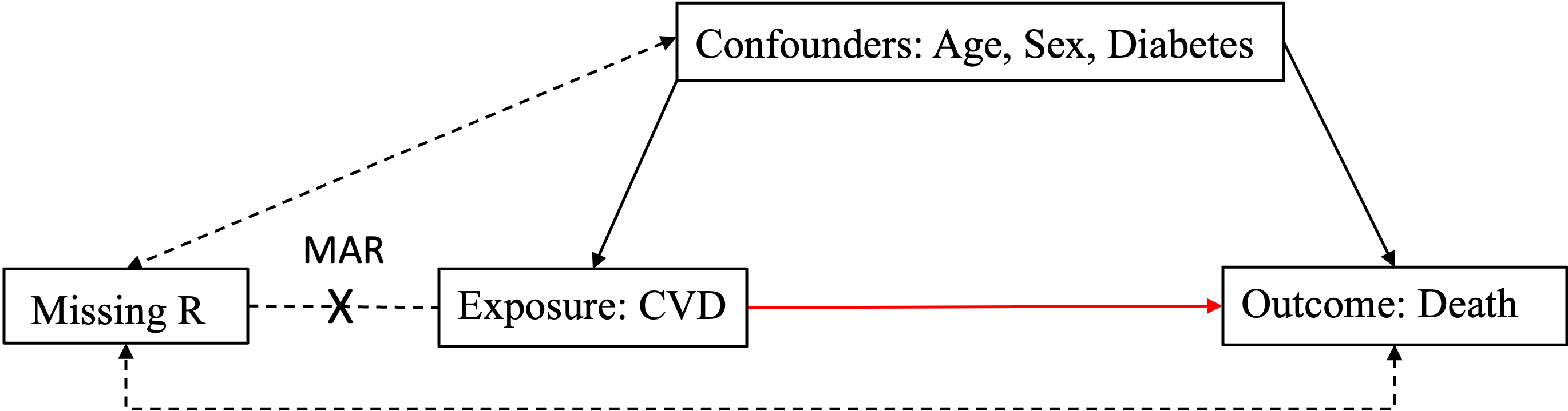}
\caption{Causal diagram for COVID-19 data as the motivating example. The black solid arrows refer to the causal relationship among confounders, the exposure, and the outcome. The double-sided dash arrows refer to the associational relationship among the missing indicator, the outcome, and the covariates. The dashed short line refers to no association between the missing indicator and the exposure given covariates and the outcome, due to MAR assumption.}
\label{fig:causal_apply}
\end{figure}

Additionally, previous clinical papers show that age is an important risk factor for CVD and death \cite{blagosklonny2020causes,lakatta2002age,wei1992age}. For COVID-19 death and coronary heart disease, sex hormones may also explain the increased risk in men vs women
\cite{dana2020insight,bots2017sex}. Moreover, severe diabetes is considered one of the reasons for death, and researchers also find the diabetes difference for patients with CVD vs those without CVD, which can be treated as a risk factor for CVD \cite{abdi2020diabetes,sowers1999diabetes}. We will assume SITA assumption $(Y^1, Y^0) \bot \text{CVD}|(\text{age, sex, diabetes})$, which includes age, sex, and diabetes as major confounders. Based on MAR, SITA assumptions, and background knowledge from clinical papers, we assume a causal diagram for COVID-19 data, as shown in Figure \ref{fig:causal_apply}.

\subsection{Estimation for Causal Effect}



Researchers may often think that CVD status causes a higher risk of death based on the common sense, but our analysis does not support this opinion. From Table \ref{table:apply}, after adjusting for both missing and confounding issues, using IPW, DR-MICE, and TR methods, their results are close to each other with estimated causal odds ratio around 0.95.  Meanwhile, we do not find a statistically significant impact of CVD on mortality because the estimated CIs for the causal odds ratio will cover 1. 

One possible reason is that higher age is not only associated with a larger risk of having CVD status but also related to a higher risk of death among those COVID-19 patients. Therefore, after balancing age and other confounders, CVD will not be the major factor causing death in this study. These results are also consistent with a recent clinical paper from \citet{vasbinder2022relationship}, which states that ``CVD risk factors, rather than CVD itself, were the major contributors to outcomes".

In conclusion, the causal odds ratio of CVD on mortality of COVID-19 via TR-WEE estimator is 0.951 with 95\% CI [0.666,1.420] based on bootstrap percentile approach.

\begin{table}[ht]
\centering
\caption{Estimated causal effect of CVD on mortality in COVID-19 data}
\label{table:apply}
\begin{tabular}{lcc}
  \hline
Method & Estimate & 95\% CI Percentile \\ 
  \hline
IPW-DR & 0.943 & (0.654,1.416) \\ 
  IPW-WEE & 0.949 & (0.661,1.418) \\ 
  DR-MICE & 0.989 & (0.706,1.480) \\ 
  TR-AIPW & 0.956 & (0.666,1.446) \\ 
  TR-WEE & 0.951 & (0.666,1.420) \\ 
   \hline
\end{tabular}
\end{table}

\section{Discussion}

When there exists missing exposure in observational studies, it is challenging to identify the causal effect of the exposure on disease in epidemiological and biomedical studies. In such cases, we have to adopt some approaches, such as inverse probability weighting, to deal with both missing and confounding issues.

The major innovation of this article is to propose WEE methods to reduce the bias and variance of estimated causal effects affected by the joint effect of extreme weights in finite samples compared with some estimators in the previous literature, such as IPW-IPW or IPW-DR \cite{zhang}. To further protect against the misspecification of the missingness or the imputation model, two TR estimators are proposed with more robust properties, which only require ``two out of three groups of models to be correct", and TR-WEE can also be more resistant to the extreme weights than TR-AIPW in finite samples.  We also provide inference results based on the bootstrap percentile, which shows that IPW-WEE or TR-WEE leads to the coverage rates of CI closer to 95\% than the other approaches.


In this paper, we also discuss the performance of imputation-based methods versus inverse weighting-based methods. Our simulation study shows that imputation-based methods may not always be valid especially when the imputation model is wrong. In contrast,  two TR estimators can protect against the wrong imputation model. 


One of the limitations for TR estimators is that when both missingness and imputation models are wrong, we require both PS and outcome models to be correct so that the Bayesian approach can be applied to consistently estimate the imputation model. However, in application studies, researchers cannot always know which model is specified correctly without enough background information. In such cases, we can still apply a Bayesian approach to estimate $P(A=1|\bm{X},Y)$ and compare with the results from the fitted imputation model. If their results are very different, then at least one of the imputations, PS, or the outcome model should be wrong. In addition, we also suggest researchers conduct a sensitivity analysis to check the model assumption or unmeasured confounders if the background knowledge is limited in real application studies \cite{ding2016sensitivity,vanderweele2017sensitivity}.

Further research can expand the IPW and TR estimators into multiple areas such as MAR on both the exposure of interest and covariates. One of ways for that problem is to impute those missing covariates and then the proposed WEE method can be easily applied to deal with missing and confounding issues. Additionally, we can also consider other missing mechanisms such as MNAR. Finally, it may be interesting for researchers to incorporate MAR or MNAR assumptions into other complex settings, such as longitudinal or survival data in future studies.

\backmatter

\bmhead{Supplementary information}

The supplementary material is available with this paper at the website. The R package {\tt{\href{https://github.com/yuliang-shi/trme }{``trme"}}}  is uploaded on GitHub and will be available for users after the publication. 

\bmhead{Acknowledgements}

We would like to thank Professor Glen McGee for giving valuable advice on the manuscript. We also appreciate Wei Liang for his great suggestions on the supplemental material. We appreciate IntegraSUS \footnotemark \footnotetext{\url{https://integrasus.saude.ce.gov.br/}} for publicly sharing the COVID-19 data in Brazil. 

\section*{Declarations}

\noindent {\it{The authors have declared no conflict of interest.}}

\begin{appendices}

\section{Appendix}\label{secA1}
\label{sec:appendix}

\subsection{Boxplot for IPW Method}

\begin{figure}[ht]
\centering
\includegraphics[width=\textwidth]{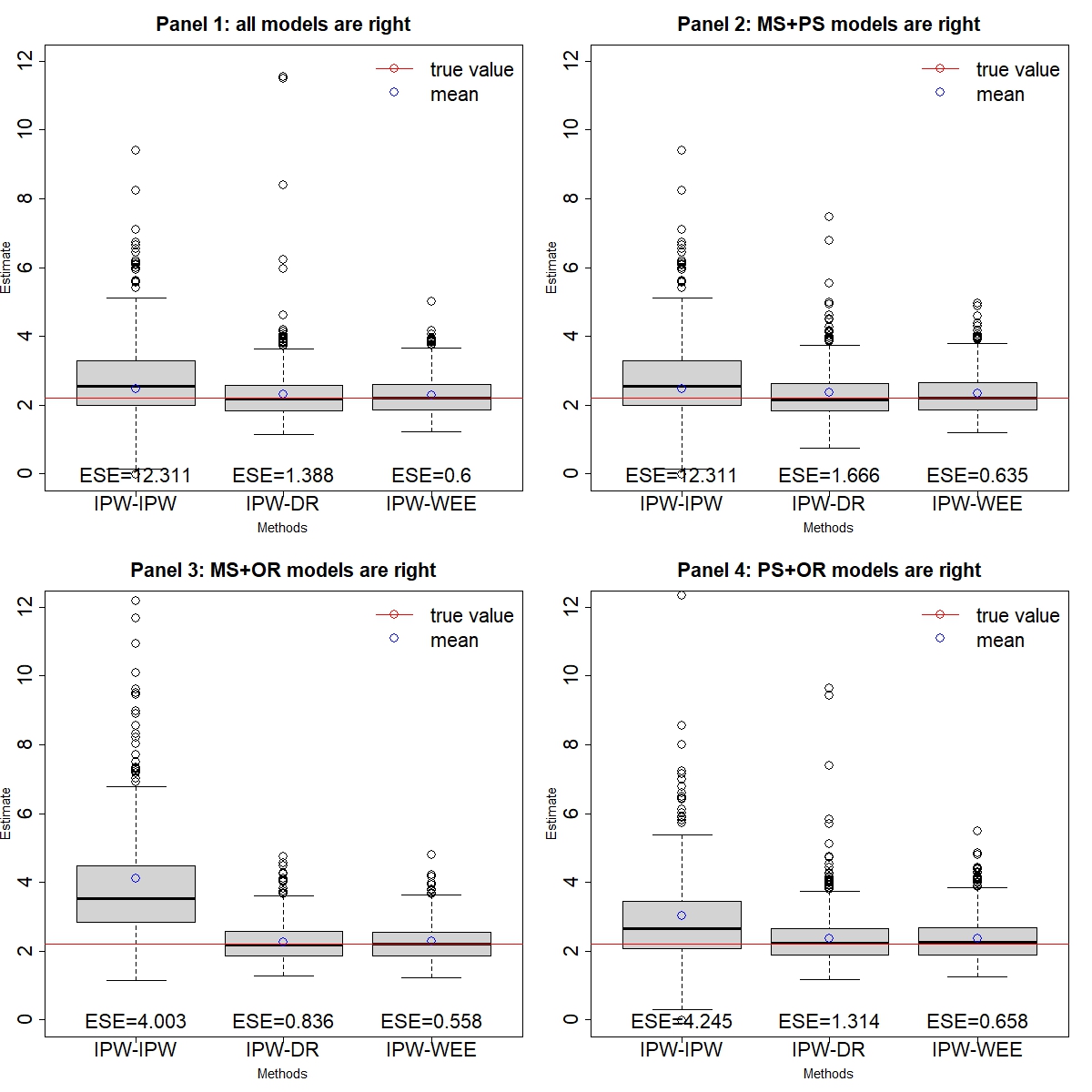}
\caption{Boxplot of point estimation using IPW methods when both missingness and PS models are correct, but the outcome model is wrong with 40\% missing rate. The true causal effect $\tau= 2.247$. The outliers are points below $\text{Q}_1-1.5 \text{IQR}$ or above $\text{Q}_3+1.5\text{IQR}$, where $\text{Q}_1$, $\text{Q}_3$, $\text{IQR}$ refer to the first quantile, the third quantile, and the interquartile range, respectively.
}
\label{fig:boxplot_ipw_sim} 
\end{figure}

The boxplot of point estimation using IPW method is provided in Figure \ref{fig:boxplot_ipw_sim}, which shows that IPW-WEE has smaller variation and ESE than IPW-DR. In contrast, IPW-IPW estimator creates some large outliers, which largely affect the point estimation and standard errors.

\subsection{Boxplot for DR and TR Methods}

\begin{figure}[ht]
\centering
\includegraphics[width=1\textwidth]{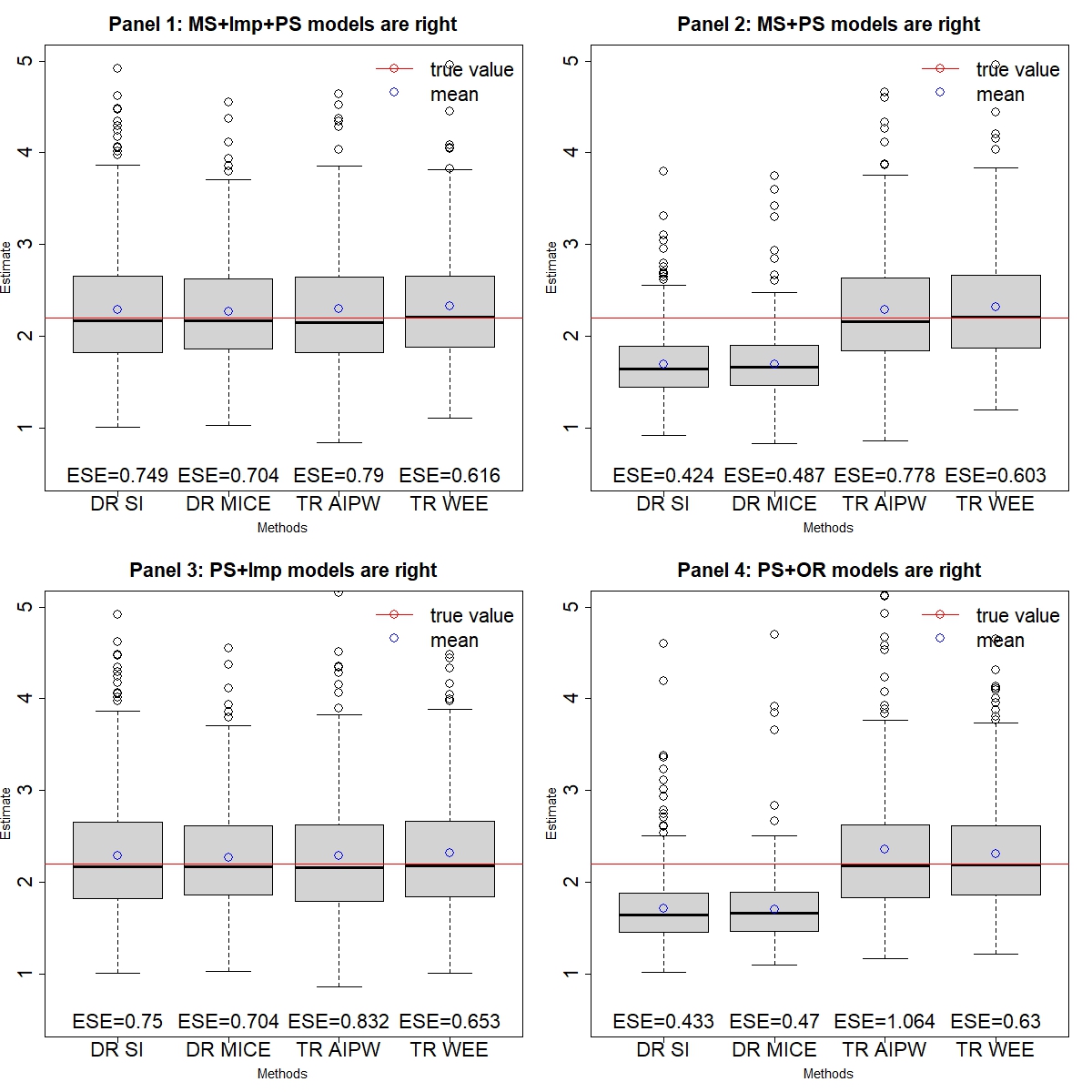}
\caption{Boxplot of point estimation using DR and TR methods when PS and outcome models are correct, but the missingness and imputation models are wrong with 40\% missing rates. The true causal effect $\tau= 2.247$. The outliers are points below $\text{Q}_1-1.5 \text{IQR}$ or above $\text{Q}_3+1.5\text{IQR}$, where $\text{Q}_1$, $\text{Q}_3$, $\text{IQR}$ refer to the first quantile, the third quantile, and the interquartile range, respectively.
}
\label{fig:boxplot_tr_sim} 
\end{figure}

The boxplot of point estimation using TR method is provided in Figure \ref{fig:boxplot_tr_sim}, which shows that TR-WEE has a smaller variation and a more narrow interquartile range (IQR) than TR-AIPW. TR-AIPW may still be affected by the extreme weights, which cause larger ESE than TR-WEE. Moreover,  both DR-SI and DR-MICE generate huge bias compared with two TR estimators when the imputation model is wrongly specified. 


\end{appendices}

\bibliography{manuscript.bib}


\renewcommand{\thesection}{S.\arabic{section}}
\renewcommand{\thesubsection}{S.\arabic{section}.\arabic{subsection}}
\renewcommand{\thetable}{S.\arabic{table}}
\renewcommand{\thefigure}{S.\arabic{figure}}
\renewcommand{\theequation}{S.\arabic{equation}}

\section{Supporting Material for Causal Inference on Missing Exposure via Robust Estimation}


\section{Asymptotic Properties of $\hat{\bm{\alpha}}^{(EE)}$ and $\hat{\bm{\beta}}^{(EE)}$}
\label{sec_proof_consist}

In this section, we prove properties of $\hat{\bm{\alpha}}^{(EE)}$ as an example. Notice that estimating both $\hat{\bm{\alpha}}^{(EE)}$ and $\hat{\bm{\beta}}^{(EE)}$ are not our main interest because they are just used as plugged-in values for IPW or TR estimators.  Our main interest is to estimate the causal effect of the exposure on the outcome, defined as $\tau=\frac{\tau_1/(1-\tau_1)}{\tau_0/(1-\tau_0)}.$ 

\subsection{Robust Estimator for PS Model}

To adjust for missingness, $\hat{\bm{\alpha}}^{(EE)}$ is constructed by the missingness and PS models, which can be written as:
\begin{equation}
\begin{aligned}
\label{alpha_DR}
S({\bm{\alpha}}^{(EE)})=\sum_{i=1}^n \left\{\frac{1-R_i}{1-P_{R_i}(\bm{\hat{\gamma}})}U_i(A_i,\bm{X}_i; \bm{\alpha})-\frac{P_{R_i}(\bm{\hat{\gamma}})-R_i}{1-P_{R_i}(\bm{\hat{\gamma}})} \hat{m}_A (\bm{X}_i,Y_i) \right\}=0,
\end{aligned}    
\end{equation}
where $U_i(A_i,\bm{X}_i; \bm{\alpha})=\frac{A_i \bm{X}_i-(1-A_i)\bm{X}_i\text{exp}(\bm{X}_i^T\bm{\alpha})}{1+\text{\text{exp}}(\bm{X}_i^T\bm{\alpha})}$ is  $(p+1) \times 1$ vector of the original score function for PS model and $\hat{m}_A (\bm{X}_i,Y_i)=E[U_i(A_i,\bm{X}_i; \bm{\alpha}) |\bm{X,Y;\hat{\delta}}]$ is the conditional expectation taken w.r.t $(A|\bm{X},Y)$. 

To prove the property of $\hat{\bm{\alpha}}^{(EE)}$, let us discuss the situation in three cases. First, if both the missingness and PS models are correct ($\gamma^*,\alpha^*$ will be true values), but the imputation model is wrong ($\widetilde{\delta}$ will be wrong value), we have $E(R_i|\bm{X,Y})=P(R_i=1|\bm{X,Y})=P_{R_i}(\bm{\gamma^*})$. Then, under the value of $(\bm{\gamma^*,\alpha^*,\widetilde{\delta}})$, Equation (\ref{alpha_DR}) will be simplified as:
\begin{equation}
\begin{aligned}
\label{alpha_DR_case1}
E[S({\bm{\alpha}}^{(EE)})]
=& E\left\{ \sum_{i=1}^n E \left[\frac{1-R_i}{1-P_{R_i}(\bm{\gamma^*})}U_i(A_i,{\bm{X}_i};{\bm{\alpha^*}})|\bm{X,Y} \right] -\sum_{i=1}^n \frac{E(R_i|\bm{X,Y})-P_{R_i}(\bm{\hat{\gamma}})}{1-P_{R_i}(\bm{\hat{\gamma}})} {m}_A(\mathbf{X,Y};\bm{\widetilde{\delta}}) \right\} \\
=&E\left[\sum_{i=1}^n\frac{1}{1-P_{R_i}(\bm{\gamma^*})}E(1-R_i|\bm{X,Y}) \times E[U_i(A_i,{\bm{X}_i};{\bm{\alpha^*}})|\bm{X,Y}]-0 \right] \\
=& E\left\{ E \left[ \sum_{i=1}^n U_i(A_i,{\bm{X}_i};{\bm{\alpha^*}}) \small|\bm{X,Y}\right] \right\} \\
=&E \left[ \sum_{i=1}^n U_i(A_i,{\bm{X}_i};{\bm{\alpha^*}}) \right]=0. 
\end{aligned}
\end{equation}
The first equality is due to the double conditional expectation on both $(\bm{X},Y_i)$. The second and third equalities are due to the MAR assumption and correct missingness model. The fourth equality is applying double expectation conditioning again.  The final equality is due to the correct PS model with the zero expectation of score function, so we will have $E \left[ \sum_{i=1}^n U_i(A_i,{\bm{X}_i};{\bm{\alpha^*}}) \right] =0$. Therefore, we know that $\hat{{\bm{\alpha}}}^{(EE)} \xrightarrow[]{\text{p}} {\bm{\alpha}^*}$ and its asymptotically normality holds as $n \to \infty$ based on the properties of estimating equations.

Next, if the missingness model is wrong ($\widetilde{\gamma}$ will be the wrong value), but the imputation model and PS model are correct ($\alpha^*,\delta^*$ will be true values), we know $E(R_i|\bm{X,Y}) \neq P_{R_i}(\bm{\gamma^*})$, but $E[U(\bm{A}|\bm{X}_i,\bm{\alpha^*})|\bm{X,Y}] = m_A(\mathbf{X,Y};\bm{\delta^*})$. Following the similar technique, under the values of $(\bm{\alpha^*,\delta^*,\widetilde{\gamma}})$, the equation will be rewritten as:
\begin{equation}
\small
\begin{aligned}
\label{alpha_DR_case2}
&E[S({\bm{\alpha}}^{(EE)})] \\
&= E \left\{\sum_{i=1}^n \left[\frac{(1-R_i)U_i(A_i,\bm{X}_i; \bm{\alpha^*})}{1-P_{R_i}(\bm{\widetilde{\gamma}})}-\frac{(1-R_i)m_A(\mathbf{X,Y};\bm{\delta^*})}{1-P_{R_i}(\bm{\widetilde{\gamma}})} \right] \right\} + E \left[\sum_{i=1}^n m_A(\mathbf{X,Y};\bm{\delta^*})  \right] \\
&= E \left\{\sum_{i=1}^n E\left\{\frac{1-R_i}{1-P_{R_i}(\bm{\widetilde{\gamma}})} [U_i(A_i,\bm{X}_i; \bm{\alpha^*})-m_A(\mathbf{X,Y};\bm{\delta^*})] \Big|\bm{X,Y} \right\} \right\} + E \left[ \sum_{i=1}^n E[U_i(A_i,\bm{X}_i; \bm{\alpha^*})|\bm{X,Y,\delta^*})]  \right] \\
&= E \left\{\sum_{i=1}^n \frac{E(1-R_i|\bm{X,Y})}{1-P_{R_i}(\bm{\widetilde{\gamma}})} \left\{E[U_i(A_i,\bm{X}_i; \bm{\alpha^*})|\bm{X,Y}]-E[U_i(A_i,\bm{X}_i; \bm{\alpha^*})|\bm{X,Y,\delta^*}] \right\}  \right\} \\
& \indent + E\left[ \sum_{i=1}^n U_i(A_i,\bm{X}_i; \bm{\alpha^*})  \right] = 0,
\end{aligned}   
\normalsize
\end{equation}
where the first equality is due to rearranging the equation into two parts. The second equality is because we take double expectation on the first part conditioning on $(\bm{X},Y_i)$ and we plug in the form of $m_A(\mathbf{X,Y};\bm{\delta^*})$. The third equality is due to MAR assumption. For the last equality, when PS and imputation models are correct, we will have $E\left[ \sum_{i=1}^n U_i(A_i,\bm{X}_i; \bm{\alpha^*}) \right]=0$ and also $E[U_i(A_i,\bm{X}_i; \bm{\alpha^*})|\bm{X,Y}]-E[U_i(A_i,\bm{X}_i; \bm{\alpha^*})|\bm{X,Y,\delta^*}]=0$.  Since we have proved $E[S({\bm{\alpha}}^{(EE)})]=0$, based on the properties of estimating equations, under the regularity condition \cite{tsiatis2007semiparametric},  $\hat{\bm{\alpha}}^{(EE)} \xrightarrow[]{\text{p}} \bm{\alpha^*}$ and its asymptotically normality holds as $n \to \infty$.

\subsection{Bayes Rule}

We continue the proof in this subsection. In the third case, when missing models and imputation models are wrong, but PS and outcome models are correct,  to consistently estimate $P(A=1|\bm{X},Y)$, we can choose an alternative way to  transform $A|\bm{X},Y$ into $A|\bm{X}$ and $Y|A,\bm{X}$ via Bayes rule:
\begin{equation}
\begin{aligned}
\label{bayes_imp}
P(A=1|\bm{X},Y=y)=\frac{P(Y=y,A=1,\bm{X})}{P(Y=y,\bm{X})} =\frac{P(Y=y|A=1,\bm{X}) P(A=1|\bm{X})}{P(Y=y|\bm{X})} 
\end{aligned}    
\end{equation}


Equation (\ref{bayes_imp}) clearly connects PS, imputation, and outcome model together. If PS and outcome models are correct, but the imputation and missingness models are wrong, we can consistently estimate $P(A=1|\bm{X},Y)$. Then, we transform the third case back to the previously solved second case. 




In conclusion, for $\hat{{\bm{\alpha}}}^{(EE)}$, if PS model is correct, and we can correctly specify either the missingness or the imputation, or the outcome model is correct, we know $\hat{{\bm{\alpha}}}^{(EE)} \xrightarrow[]{\text{p}} {\bm{\alpha}^*}$ and its asymptotic normality holds as $n \to \infty$, based on the properties of estimating equations.

\subsection{Robust Estimator for Outcome Model}
\label{sec_beta_DR_miss}
Similarly, we can construct the estimator $\hat{\bm{\beta}}^{(EE)}$ based on the missingness model and the outcome model, written as:
\begin{equation}
\begin{aligned}
\label{beta_DR_Miss}
S({\bm{\beta}}^{(EE)})=\sum_{i=1}^n \left\{\frac{(1-R_i)V_{i}(Y_i,A_i, \bm{X}_i; \bm{\beta})}{1-P_{R_i}(\bm{\hat{\gamma}})}-\frac{P_{R_i}(\bm{\hat{\gamma}})-R_i}{1-P_{R_i}(\bm{\hat{\gamma}}) } \hat{m}_Y (\bm{X}_i,Y_i)\right\}=0,
\end{aligned}    
\end{equation}
where $V_{i}(Y_i,A_i, \bm{X}_i; \bm{\beta})=\frac{Y_i \bm{X}_i-(1-Y_i)\bm{X}_i\text{exp}(\bm{X}_i^T\bm{\beta}+A_i\beta_A)}{1+\text{exp}(\bm{X}_i^T\bm{\beta}+A_i\beta_A)}$ is the $(p+2) \times 1$ vector of the original score function from the outcome model. $\hat{m}_{Y} (\bm{X}_i,Y_i) =E[V_{i}(Y_i,A_i, \bm{X}_i; \bm{\beta}) |\bm{X,Y;\hat{\delta}}]$ is conditional expectation taken w.r.t $A|\bm{X},Y$ after we plug in $\hat{\delta}$. 

Following similar steps from Equation (\ref{alpha_DR_case1}-\ref{bayes_imp}), we can easily prove that if the outcome model requires to be correct and either the missingness, PS or the imputation model is also correct, we know  $\hat{\bm{\beta}}^{(EE)} \xrightarrow[]{\text{p}}  \bm{\beta}^*$, and its asymptotic normality holds as $n \to \infty$.  Again, estimating the coefficient of the outcome model is just used as plugged-in values, which is not our main interest.


\section{IPW Approaches}
\label{sec_IPW}

Our main goal is to estimate the true causal effect $\tau$ after adjusting for both missing and confounding issues.  This section discusses the newly developed IPW-WEE estimator and reviews the two types of IPW estimators from \citet{zhang}.

\subsection{Asymptotic Properties of IPW-WEE Estimator}

Both IPW estimators may be affected by extreme weights in finite samples. In comparison, IPW-WEE is developed to be more resistant to extreme weights in finite samples. In this subsection, we want to prove the asymptotic properties of IPW-WEE and clarify the connection with IPW-DR estimator. 

In the main manuscript, we have described the two steps to obtain IPW-WEE estimator. First, we revise Equation (\ref{beta_DR_Miss}), and denote the new estimate of outcome coefficients as $\hat{\bm{\beta}}_1^{(IPW-WEE)}$ solving the equation for the exposure group: 
\begin{equation}
\begin{aligned}
\label{beta,IPW-WEE}
S(\bm{\beta}_1^{(IPW-WEE)})=\sum_{i=1}^{n}  \frac{1-R_i}{1-P_{R_i}(\bm{\hat{\gamma}})} \bm{X}_i \left\{ \frac{A_i}{P_{A_i}(\bm{\hat{\alpha}}^{(WLA)})}[Y_i-E(Y_i|A_i=1,\bm{X}_i;\bm{\beta}_1)]\right\}=0,
\end{aligned}
\end{equation}
where the subscript ``1" refers to the exposure group and ``0" refers to the control group. Here, $\bm{X}_i$ requires to contain $p$ covariates and one intercept. $E(Y_i|A_i=1,\bm{X}_i;\bm{\beta}_1)=\text{expit}(\bm{X}_i^T\bm{\beta}_1)$ is the outcome model, which contains unknown $\bm{\beta}$ to be solved.

Second, we estimate $\tau_1$ by a IPW estimator using WEE, denoted as $\hat{\tau}_1^{(IPW-WEE)}$:
\begin{equation}
\begin{aligned}
\label{IPW-WEE}
\hat{\tau}_1^{(IPW-WEE)}=
&\frac{1}{n} \sum_{i=1}^n \frac{1-R_i}{1-P_{R_i}({\bm{\hat{\gamma}}})} E \left[Y_i|A_i=1,\bm{X}_i,\hat{\bm{\beta}}_1^{(IPW-WEE)} \right],
\end{aligned}
\end{equation}

Similarly, for the control group, we simply rewrite the equations as: 
\begin{equation}
\begin{aligned}
\label{beta0,IPW-WEE}
&S(\bm{\beta}_0^{(IPW-WEE)})=\sum_{i=1}^{n}  \frac{1-R_i}{1-P_{R_i}(\bm{\hat{\gamma}})} \bm{X}_i \left\{ \frac{1-A_i}{1-P_{A_i}(\bm{\hat{\alpha}}^{(WLA)})}[Y_i-E(Y_i|A_i=0,\bm{X}_i;\bm{\beta}_0)]\right\}=0, \\
&\hat{\tau}_0^{(IPW-WEE)}=
\frac{1}{n} \sum_{i=1}^n \frac{1-R_i}{1-P_{R_i}({\bm{\hat{\gamma}}})} E \left[Y_i|A_i=0,\bm{X}_i,\hat{\bm{\beta}}_0^{(IPW-WEE)} \right],
\end{aligned}
\end{equation}

In the next part, we want to prove $\hat{\bm{\beta}}_1^{(IPW-WEE)} \xrightarrow[]{\text{p}} \bm{\beta}_1^*$ and its asymptotic normality holds as $n \to \infty$. When the missingness and outcome models are correct, under the true value of $(\bm{\alpha^*,\beta^*_1})$, we take expectation on the estimating equations: 
\begin{equation}
\begin{aligned}
\label{beta,IPW-WEE,proof}
E[S(\bm{\beta}_1^{(IPW-WEE)})]
&=  \sum_{i=1}^{n} E\left\{ E\left\{ \frac{1-R_i}{1-P_{R_i}(\bm{\hat{\gamma}})} \bm{X}_i \left[ \frac{A_i}{P_{A_i}(\bm{\alpha^*})}[Y_i-E(Y_i|A_i=1,\bm{X}_i;\bm{\beta}_1)]\right] \Bigg|\bm{X,Y} \right\}\right\}\\
&=  \sum_{i=1}^{n} E\left\{ E\left[ \frac{1-R_i}{1-P_{R_i}(\bm{\hat{\gamma}})}  \Bigg|\bm{X,Y} \right]E\left\{ \bm{X}_i \left[ \frac{A_i}{P_{A_i}(\bm{\alpha^*})}[Y_i-E(Y_i|A_i=1,\bm{X}_i;\bm{\beta}_1)]\right] \Bigg|\bm{X,Y} \right\}  \right\}\\
&= \sum_{i=1}^{n} E\left\{  \frac{A_i}{P_{A_i}(\bm{\alpha^*})} \bm{X}_i[Y_i-E(Y_i|A_i=1,\bm{X}_i;\bm{\beta}_1)]  \right\} \\
&= \sum_{i=1}^{n} E\left\{ E\left\{ \frac{A_i}{P_{A_i}(\bm{\alpha^*})} \bm{X}_i[Y_i-E(Y_i|A_i=1,\bm{X}_i;\bm{\beta}_1)] \Bigg|\bm{A,X} \right\} \right\} \\
&=\sum_{i=1}^{n} E\left\{  \frac{A_i}{P_{A_i}(\bm{\alpha^*})} \bm{X}_i \left\{ E[Y_i|A_i=1,\bm{X}]- E(Y_i|A_i=1,\bm{X}_i;\bm{\beta}_1)]  \right\}\right\}=0
\end{aligned}
\end{equation}
where the first equality takes double expectation condition on $(\bm{X},Y_i)$. The second and the third equalities are due to MAR assumption and the correct missingness model. The forth equality is due to the double expectation conditioning on $(\bm{A,X})$. In the fifth equality, we know that $E[A_iY_i|\bm{A,X}]=E[A_iY_i^1|\bm{A,X}]=A_iE[Y_i^1|A_i=1,\bm{X}]=A_iE[Y_i|A_i=1,\bm{X}]$ due to the definition of $Y_i=A_iY_i^1+(1-A_i)Y_i^0$ and SITA assumption. The last equality is due to the correct outcome model, so we know $E[Y_i|A_i=1,\bm{X}]- E(Y_i|A_i=1,\bm{X}_i;\bm{\beta}_1)=0$. Based on the properties of estimating equations, since $E[S(\bm{\beta}_1^{(IPW-WEE)})]=0$, under the common regularity conditions, we know $\hat{\bm{\beta}}_1^{(IPW-WEE)} \xrightarrow[]{\text{p}} \bm{\beta}_1^*$ and its asymptotic normality holds as $n \to \infty$. Applying the similar steps, the same properties will also hold for $\hat{\bm{\beta}}_0^{(IPW-WEE)}$.

\subsection{Connection between IPW-WEE and IPW-DR Estimators}

In this part, we will prove that IPW-WEE estimator can be exactly transformed into another IPW-DR estimator after plugging in $\hat{\bm{\beta}}_1^{(IPW-WEE)}$, which is denoted as $\hat{\tau}_1^{(IPW-DR)}(\hat{\bm{\beta}}_1^{(IPW-WEE)})$ and written as:
\begin{equation}
\begin{aligned}
\label{IPW-DR,WEE}
&\hat{\tau}_1^{(IPW-DR)}(\hat{\bm{\beta}}_1^{(IPW-WEE)}) \\
&=\frac{1}{n}\sum_{i=1}^n \frac{1-R_i}{1-P_{R_i}(\bm{\hat{\gamma}})} \left\{ \frac{A_i}{P_{A_i}(\bm{\hat{\alpha}}^{(WLA)})}Y_i-\frac{A_i-P_{A_i}(\bm{\hat{\alpha}}^{(WLA)})}{P_{A_i}(\bm{\hat{\alpha}}^{(WLA)})}E(Y_i|A_i=1,\bm{X}_i;\hat{\bm{\beta}}_1^{(IPW-WEE)})  \right\} \\
&=\frac{1}{n}\sum_{i=1}^{n}  \frac{1-R_i}{1-P_{R_i}(\bm{\hat{\gamma}})} \left\{ \frac{A_i}{P_{A_i}(\bm{\hat{\alpha}}^{(WLA)})}[Y_i-E(Y_i|A_i=1,\bm{X}_i;\hat{\bm{\beta}}_1^{(IPW-WEE)})]\right\}\\
& \indent +\frac{1}{n} \sum_{i=1}^n \frac{1-R_i}{1-P_{R_i}({\bm{\hat{\gamma}}})} E \left[Y_i|A_i=1,\bm{X}_i,\hat{\bm{\beta}}_1^{(IPW-WEE)} \right] \\
&=\frac{1}{n}S_0(\hat{\bm{\beta}}_1^{(IPW-WEE)})+\hat{\tau}_1^{(IPW-WEE)} \\
&=0+\hat{\tau}_1^{(IPW-WEE)}
\end{aligned}    
\end{equation}
where $S_0(\hat{\bm{\beta}}_1^{(IPW-WEE)})$ is the first equation from Equation (\ref{beta,IPW-WEE}) when $\bm{X}_i$ uses an intercept. The first equality is to simply plug $\hat{\bm{\beta}}_1^{(IPW-WEE)}$ into the IPW-DR estimator. The second and third equalities is to rearrange terms as $S_0(\hat{\bm{\beta}}_1^{(IPW-WEE)})$ and $\hat{\tau}_1^{(IPW-WEE)}$. Since $\bm{X}_i$ contains an intercept and $\hat{\bm{\beta}}_1^{(IPW-WEE)}$ is the solution of Equation (\ref{beta,IPW-WEE}), we know $S_0(\hat{\bm{\beta}}_1^{(IPW-WEE)})=0$. Therefore,  we know that two IPW estimators will be exactly same, written as: $\hat{\tau}_1^{(IPW-WEE)}=\hat{\tau}_1^{(IPW-DR)}(\hat{\bm{\beta}}_1^{(IPW-WEE)})$. 

Finally,  when both missingness and outcome models are correct, we know $\hat{\bm{\beta}}_1^{(IPW-WEE)} \xrightarrow[]{\text{p}} \bm{\beta}_1^*$, so it can replace the role of $\hat{\bm{\beta}}_1^{(WLA)}$ as plugged-in values, also described in the main manuscript. Therefore, if the missingness model is correct and either PS or outcome model is correct, we know $\hat{\tau}_1^{(IPW-DR)}(\hat{\bm{\beta}}_1^{(IPW-WEE)})$ can achieve asymptotic consistency and normality. 

Because IPW-WEE is exactly same as IPW-DR after plugging in $\hat{\bm{\beta}}_1^{(IPW-WEE)}$, $\hat{\tau}_1^{(IPW-WEE)} \xrightarrow[]{\text{p}} {\tau_1}$ and its asymptotic normality holds as $n \to \infty$. Similarly, $\hat{\tau}_0^{(IPW-WEE)}$ is denoted as IPW-WEE estimator for $\tau_0$ after replacing $A_i$ with $1-A_i$.  Then, we know $\hat{\tau}^{(IPW-WEE)}= \frac{\hat{\tau}_1^{(IPW-WEE)}/(1-\hat{\tau}_1^{(IPW-WEE)})}{\hat{\tau}_0^{(IPW-WEE)}/(1-\hat{\tau}_0^{(IPW-WEE)})} \xrightarrow[]{\text{p}} \tau$. 


The major advantage of $\hat{\tau}_1^{(IPW-WEE)}$ is to reduce the effect of extreme weights in finite samples, compared with the previous IPW-DR estimator from \citet{zhang} because $\hat{\bm{\beta}}_1^{(IPW-WEE)}$ absorbs the joint effect of extreme weights of the missingness and PS models. After that, $\hat{\tau}_1^{(IPW-WEE)}$ will only involve inverse weights of missingness without weights of PS values. Therefore, the new IPW-WEE outperforms other IPW estimators in the simulation study.

\section{Theory of TR Properties}
\label{sec_tr}

\subsection{Consistency for TR-AIPW Estimator}

In this section, we want to prove that TR estimator using AIPW method contain TR group properties, i.e. it only requires ``two out of three groups of models to be correct" to achieve consistency.  To construct TR estimators, we require more robust estimators as plugged-in values in Equation (\ref{alpha_DR}) because now the missingness model may not be always correct. Therefore, for all TR estimators, we will use $(\bm{\hat{\alpha}^{(EE)},\hat{\beta}^{(EE)}})$ instead of $(\bm{\hat{\alpha}}^{(WLA)},\bm{\hat{\beta}}^{(WLA)})$ compared with IPW estimators.

TR-AIPW estimator for $\tau_1$ will be the combination of two DR estimators to adjust for both missing and confounding issues, and we denote the TR estimator using augmented inverse probability weighting (AIPW) as $\hat{\tau}_1^{(TR-AIPW)}$:
\begin{equation}
\begin{aligned}
\label{TR-AIPW}
\hat{\tau}_1^{(TR-AIPW)}
&=\frac{1}{n} \sum_{i=1}^n \frac{1-R_i}{1-P_{R_i}({\bm{\hat{\gamma}}})} \bm{Q_{i1}}(Y_i,A_i,\bm{X_i;\hat{\alpha}^{(EE)},\hat{\beta}^{(EE)}}_1 )  \\
& \indent -\frac{1}{n} \sum_{i=1}^n \frac{P_{R_i}({\bm{\hat{\gamma}}})-R_i}{1-P_{R_i}({\bm{\hat{\gamma}}})}E\left[ \bm{Q_{i1}}(Y_i,A_i,\bm{X_i;\hat{\alpha}^{(EE)},\hat{\beta}^{(EE)}}_1 )  \Bigg |\bm{X,Y;\hat{\delta}}\right] ,
\end{aligned}
\end{equation}
where  
$\bm{Q_{i1}}(Y_i,A_i,\bm{X_i;\hat{\alpha}^{(EE)},\hat{\beta}^{(EE)}})= \frac{A_i}{P_{A_i}(\hat{\bm{\alpha}}^{(EE)})}Y_i-\frac{A_i-P_{A_i}(\hat{\bm{\alpha}}^{(EE)})}{P_{A_i}(\hat{\bm{\alpha}}^{(EE)})}E(Y_i|A_i=1,\bm{X}_i;\hat{\bm{\beta}}^{(EE)}_1)$. Here,$(\hat{\bm{\gamma}},\hat{\bm{\delta}})$ are estimated from the missingness and the imputation models, which are defined in Section 2 of the manuscript, respectively. $(\hat{\bm{\alpha}}^{(EE)},\hat{\bm{\beta}}_1^{(EE)})$ are plug-in values for PS values and fitted response for the exposure group, respectively. Both are estimated values after adjusting for the missingness from Equation (\ref{alpha_DR}-\ref{beta_DR_Miss}). Notice that $E[\bm{Q_{i1}}(Y_i,A_i,\bm{X_i;\hat{\alpha}^{(EE)},\hat{\beta}^{(EE)}}_1 ) |\bm{X,Y}]$ is taken w.r.t $A|\bm{X},Y$ after substituting $\bm{\hat{\delta}}$
into unknown $\bm{\delta}$.  

Similarly, TR-AIPW estimator for $\tau_0$ will be written as:
\begin{equation}
\begin{aligned}
\label{TR-AIPW0}
\hat{\tau}_0^{(TR-AIPW)}
&=\frac{1}{n} \sum_{i=1}^n  \frac{1-R_i}{1-P_{R_i}({\bm{\hat{\gamma}}})} \bm{Q_{i0}}(Y_i,A_i,\bm{X_i;\hat{\alpha}^{EE},\hat{\beta}^{EE}}_0)  \\
& \indent -\frac{1}{n} \sum_{i=1}^n \frac{P_{R_i}({\bm{\hat{\gamma}}})-R_i}{1-P_{R_i}({\bm{\hat{\gamma}}})}E\left[ \bm{Q_{i0}}(Y_i,A_i,\bm{X_i;\hat{\alpha}^{EE},\hat{\beta}^{EE}}_0) \Bigg |\bm{X,Y;\hat{\delta}}\right] ,
\end{aligned}
\end{equation}
where  
$\bm{Q_{i0}}(Y_i,A_i,\bm{X_i;\hat{\alpha}^{(EE)},\hat{\beta}^{(EE)}})= \frac{1-A_i}{1-P_{A_i}(\hat{\bm{\alpha}}^{(EE)})}Y_i-\frac{P_{A_i}(\hat{\bm{\alpha}}^{(EE)})-A_i}{1-P_{A_i}(\hat{\bm{\alpha}}^{(EE)})}E(Y_i|A_i=0,\bm{X}_i;\hat{\bm{\beta}}^{(EE)}_0)$. 

Now we will prove the consistency of $\hat{\tau}_1^{(TR-AIPW)}$ in three different scenarios. First, if the missingness model is correct ($\gamma^*$ is true value), the imputation model is wrong ($\widetilde{\delta}$ is wrong value), and either PS or outcome model is correct ($\alpha^*$ or $\beta_1^*$ is true value), under the value of $(\bm{\alpha^*\text{ or }\beta^*_1,\gamma^*,\widetilde{\delta}})$, we can simplify the expectation of Equation (\ref{TR-AIPW}) as:
\begin{equation}
\small
\label{TR-AIPW proof1}
\begin{aligned}
E[\tau_1^{(TR-AIPW)}] 
&=\frac{1}{n} E\left\{ \sum_{i=1}^n E \left[\frac{1-R_i}{1-P_{R_i}(\bm{\gamma^*})}\bm{Q_{i1}}(Y_i,A_i,\bm{X_i;\alpha^*\text{ or }\beta^*_1}) |\bm{X,Y} \right] \right\} \\
& \indent -\frac{1}{n} E\left\{\sum_{i=1}^n \frac{P_{R_i}(\bm{\gamma^*})-E[R_i|\bm{X,Y}]}{1-P_{R_i}(\bm{\gamma^*})}E[\bm{Q_{i1}}(Y_i,A_i,\bm{X_i;\alpha^*\text{ or }\beta^*_1})|\bm{X,Y},\bm{\widetilde{\delta}}] \right\}\\
&=\frac{1}{n} E\left\{ E\left[\sum_{i=1}^n  \frac{E[1-R_i|\bm{X,Y}]}{1-P_{R_i}(\bm{\gamma^*})} \right] E \left[ \sum_{i=1}^n \bm{Q_{i1}}(Y_i,A_i,\bm{X_i;\alpha^*\text{ or }\beta^*_1})|\bm{X,Y} \right]-0 \right\}  \\
&=\frac{1}{n} E \left\{ E\left[\sum_{i=1}^n \bm{Q_{i1}}(Y_i,A_i,\bm{X_i;\alpha^*\text{ or }\beta^*_1}) |\bm{X,Y} \right] \right\} \\
&=\frac{1}{n} E\left[\sum_{i=1}^n \bm{Q_{i1}}(Y_i,A_i,\bm{X_i;\alpha^*\text{ or }\beta^*_1}) \right]\\
&=E[Y^1]. 
\end{aligned}
\normalsize
\end{equation}
The first equality is due to the double expectation conditioning on $(\bm{X},Y_i)$. The second and third equalities are due to the MAR assumption and the correct missingness model. Since  $\bm{Q_{i1}}(Y_i,A_i,\bm{X_i;\alpha^*\text{ or }\beta^*_1})|\bm{X,Y}$ is a function of $A|\bm{X},Y$, it will be conditionally independent of $\bm{R}$ due to MAR assumption. The fourth equality holds due to the double expectation. The final equality holds because $\frac{1}{n}\sum_{i=1}^n \bm{Q_{i1}}(Y_i,A_i,\bm{X_i;\alpha^*\text{ or }\beta^*_1})$ is a DR estimator, which only requires either PS or the outcome model to be correct. 

In the second case, if the missingness model is wrong ($\widetilde{\gamma}$ is wrong value), but the imputation model is correct ($\delta^*$ are true values) and either PS or the outcome model is correct ($\alpha^*  \text{ or } \beta^*_1$ is the true value). Then, under the values of $(\bm{\widetilde{\gamma},\delta^*,\alpha^* \text{ or } \beta^*_1})$, we can rewrite Equation (\ref{TR-AIPW}) as:
\begin{equation}
\small
\label{TR-AIPW proof2}
\begin{aligned}
&E[\tau_1^{(TR-AIPW)}] \\
&=\frac{1}{n} E\left\{\sum_{i=1}^n \frac{1-R_i}{1-P_{R_i}(\bm{\widetilde{\gamma}})} \left\{ \bm{Q_{i1}}(Y_i,A_i,\bm{X_i;\alpha^*  \text{ or } \beta^*_1})-E[\bm{Q_{i1}}(Y_i,A_i,\bm{X_i;\alpha^*  \text{ or } \beta^*_1})|\bm{X,Y,\delta^*} ] \right\} \right\}  \\
&\indent + \frac{1}{n} E \left\{\sum_{i=1}^n E[\bm{Q_{i1}}(Y_i,A_i,\bm{X_i;\alpha^*  \text{ or } \beta^*_1})|\bm{X,Y,\delta^*} ] \right\} \\
&= \frac{1}{n}\sum_{i=1}^n E \left\{  \frac{1-E[R_i|\bm{X,Y}]}{1-P_{R_i}(\bm{\widetilde{\gamma}})} \left\{  E[\bm{Q_{i1}}(Y_i,A_i,\bm{X_i;\alpha^*  \text{ or } \beta^*_1})|\bm{X,Y}] -E\left[\bm{Q_{i1}}(Y_i,A_i,\bm{X_i;\alpha^*  \text{ or } \beta^*_1})|\bm{X,Y,\delta^*}  \right] \right\} \right\} \\
&\indent + \frac{1}{n} E\left[ \sum_{i=1}^n \bm{Q_{i1}}(Y_i,A_i,\bm{X_i;\alpha^*  \text{ or } \beta^*_1})\right]\\
&= 0+E[Y^1].
\end{aligned}
\normalsize
\end{equation}
The first equation rearranges the equation into two parts. The second equality is due to the double expectation condition on $(\bm{X},Y_i)$ and MAR assumption. In the third equality,  $E[\bm{Q_{i1}}(Y_i,A_i,\bm{X_i;\alpha^*  \text{ or } \beta^*_1})|\bm{X,Y}]$ is taken w.r.t $A|\bm{X},Y$. If the imputation model is correct, we know $ E[\bm{Q_{i1}}(Y_i,A_i,\bm{X_i;\alpha^*  \text{ or } \beta^*_1})|\bm{X,Y}] -E\left[\bm{Q_{i1}}(Y_i,A_i,\bm{X_i;\alpha^*  \text{ or } \beta^*_1})|\bm{X,Y,\delta^*}  \right]=0$ because both are a function of $A|X,Y$. In addition, if either PS or outcome model is correct, we know $\frac{1}{n} E\left[ \sum_{i=1}^n \bm{Q_{i1}}(Y_i,A_i,\bm{X_i;\alpha^*  \text{ or } \beta^*_1})\right]=E[Y^1]$ due to DR property. 

In the third case, when both missingness and imputation models are wrong, but both PS and outcome models are correct, we can still consistently estimate $P(A=1|\bm{X},Y)$ based on the Bayes rules from Equation (\ref{bayes_imp}). In other words, the imputation model can be correctly specified because $\delta^*$ can be consistently estimate as a function of $(\hat{\alpha},\hat{\beta})$. Then, the third case can be transformed back to the previously solved case. Following the same proof of Equation (\ref{TR-AIPW proof2}), under the true values of $(\bm{\widetilde{\gamma},\delta^*,\alpha^*,\beta^*_1})$, we know $E[\tau_1^{(TR-AIPW)}]=E[Y^1]$. Certainly, for $\tau_0=E[Y^0]$, we also prove that $E[\tau_0^{(TR-AIPW)}]=E[Y^0]$ after replace $A_i$ with $1-A_i$. 

Finally, TR-AIPW estimator for $\tau$ will also be consistent with the true value, written as:
\begin{equation}
\begin{aligned}
\hat{\tau}^{TR-AIPW}= \frac{\hat{\tau}_1^{(TR-AIPW)}/(1-\hat{\tau}_1^{(TR-AIPW)})}{\hat{\tau}_0^{(TR-AIPW)}/(1-\hat{\tau}_0^{(TR-AIPW)})} \xrightarrow[]{\text{p}}{\tau}.  
\end{aligned}    
\end{equation}
In summary, we prove TR robust properties as ``two out of three groups of models to be correct", which means if we classify the missingness and imputation models as missing/imputing model group, PS model as the second group, and the outcome model as the third group,  as long as any two groups can be correctly specified, we know $\hat{\tau}^{(TR-AIPW)} \xrightarrow[]{\text{p}} {\tau}$.

\subsection{Connection between Two TR Estimators}

From the main manuscript, we conduct WEE estimator for the coefficients of the outcome model as $\hat{\bm{\beta}}_1^{(TR-WEE)}$ to solve the following equation: 
\begin{equation}
\begin{aligned}
\small
\label{beta,TR-WEE}
S(\bm{\beta}_1^{(TR-WEE)})=  &\sum_{i=1}^{n} \frac{1-R_i}{1-P_{R_i} (\bm{\hat{\gamma}})}\frac{\bm{X}_i}{P_{A_i}(\bm{\hat{\alpha}^{(EE)}})}  \left\{ A_i\Delta_i(Y_i,\bm{X}_i;\bm{\beta}_1)- E\left[A_i \Delta_i(Y_i,\bm{X}_i;\hat{\bm{\beta}}_1^{(EE)})\Bigg |\bm{X,Y;\hat{\delta}} \right] \right\} \\
&+ \sum_{i=1}^n E\left[ \bm{Q_{i1}}(Y_i,A_i,\bm{X_i;\hat{\alpha}^{(EE)},\hat{\beta}^{(EE)}_1}) \Bigg |\bm{X,Y;\hat{\delta}}\right]
=0,
\end{aligned}
\normalsize
\end{equation}
where $\Delta(Y_i,\bm{X}_i; \hat{\bm{\beta}}_1)=Y_i-E(Y_i|A_i=1,\bm{X}_i;\bm{\beta}_1)$ and $E(Y_i|A_i=1,\bm{X}_i;\bm{\beta}_1)=\text{expit}(\bm{X}_i^T \bm{\beta}_1)$ is the outcome model with unknown $(p+1) \times 1$ vector of $\bm{\beta}_1$ when individuals are in the exposure group. Here, $\bm{Q_{i1}}(Y_i,A_i,\bm{X_i;\hat{\alpha}^{(EE)},\hat{\beta}^{(EE)}})= \frac{A_i}{P_{A_i}(\hat{\bm{\alpha}}^{(EE)})}Y_i-\frac{A_i-P_{A_i}(\hat{\bm{\alpha}}^{(EE)})}{P_{A_i}(\hat{\bm{\alpha}}^{(EE)})}E(Y_i|A_i=1,\bm{X}_i;\hat{\bm{\beta}}^{(EE)}_1)$ and $(\bm{\hat{\alpha}^{(EE)}},\bm{\hat{\beta}^{(EE)}})$ are estimated values from Equations (\ref{alpha_DR})-(\ref{beta_DR_Miss}). Here, $\bm{X}_i$  includes one intercept and $p$ covariates.

To study the connection between TR-AIPW and TR-WEE, we rewrite TR-AIPW as:
\begin{equation}
\begin{aligned}
\label{TR-Connect}
&\hat{\tau}_1^{(TR-AIPW)}(\hat{\bm{\beta}}_1^{(TR-WEE)}, \hat{\bm{\beta}}_1^{(EE)}) \\
&=\frac{1}{n}\sum_{i=1}^n \frac{1-R_i}{1-P_{R_i}(\bm{\hat{\gamma}})} \left\{ \frac{A_i}{P_{A_i}(\bm{\hat{\alpha}^{(EE)}})}Y_i-\frac{A_i-P_{A_i}(\bm{\hat{\alpha}^{(EE)}})}{P_{A_i}(\bm{\hat{\alpha}^{(EE)}})}E(Y_i|A_i=1,\bm{X}_i;\hat{\bm{\beta}}_1^{(TR-WEE)})  \right\} \\
& \indent -\frac{1}{n}\sum_{i=1}^{n}  \frac{1-R_i}{1-P_{R_i}(\bm{\hat{\gamma}})} E\left\{ \bm{Q_{i1}}(Y_i,A_i,\bm{X_i;\hat{\alpha}^{(EE)},\hat{\beta}^{(EE)}}_1 )|\bm{X,Y,\hat{\delta}}\right\} \\
& \indent +\frac{1}{n}\sum_{i=1}^{n} E\left\{ \bm{Q_{i1}}(Y_i,A_i,\bm{X_i;\hat{\alpha}^{(EE)},\hat{\beta}^{(EE)}}_1 )|\bm{X,Y,\hat{\delta}}\right\}  \\
&=\frac{1}{n}\sum_{i=1}^{n}  \frac{1-R_i}{1-P_{R_i}(\bm{\hat{\gamma}})} \frac{A_i}{P_{A_i}(\bm{\hat{\alpha}^{(EE)}})}[Y_i-E(Y_i|A_i=1,\bm{X}_i;\hat{\bm{\beta}}_1^{(TR-WEE)})] \\
& \indent - \frac{1}{n}\sum_{i=1}^{n}  \frac{1-R_i}{1-P_{R_i}(\bm{\hat{\gamma}})}E \left\{\frac{A_i}{P_{A_i}(\bm{\hat{\alpha}^{(EE)}})}[Y_i-E(Y_i|A_i=1,\bm{X}_i;\hat{\bm{\beta}}_1^{(EE)})]| \bm{X,Y,\hat{\delta}} \right\} \\
& \indent +\frac{1}{n}\sum_{i=1}^{n} \frac{1-R_i}{1-P_{R_i}(\bm{\hat{\gamma}})} [E(Y_i|A_i=1,\bm{X}_i;\hat{\bm{\beta}}_1^{(TR-WEE)})-E(Y_i|A_i=1,\bm{X}_i;\hat{\bm{\beta}}_1^{(EE)}) ]  \\
& \indent +\frac{1}{n}\sum_{i=1}^{n} E\left\{ \bm{Q_{i1}}(Y_i,A_i,\bm{X_i;\hat{\alpha}^{(EE)},\hat{\beta}^{(EE)}}_1 )|\bm{X,Y,\hat{\delta}}\right\}  \\
&=\frac{1}{n}S_0(\hat{\bm{\beta}}_1^{(TR-WEE)})+\hat{\tau}_1^{(TR-WEE)} \\
&=0+\hat{\tau}_1^{(TR-WEE)},
\end{aligned}
\normalsize
\end{equation}
where the first equality is to plug in $\hat{\bm{\beta}}_1^{(TR-WEE)}, \hat{\bm{\beta}}_1^{(EE)}$ into TR-AIPW and the second equality is to rearrange four terms and connect TR-AIPW with TR-WEE. Here, $S_0(\hat{\bm{\beta}}_1^{(TR-WEE)})$ is the first equation from the set of estimating equation (\ref{beta,TR-WEE}) when the first element of $\bm{X}_i$ is an intercept. For the third and last equality, similar as IPW-WEE estimator,  because $\hat{\bm{\beta}}_1^{(TR-WEE)}$ is the solution of Equation (\ref{beta,TR-WEE}), we know $S_0(\hat{\bm{\beta}}_1^{(TR-WEE)})=0$ and TR-WEE will be exactly same as TR-AIPW after plugging into $\hat{\bm{\beta}}_1^{(TR-WEE)}$. 

The key difference between TR-WEE and IPW-WEE is that we replace the estimated coefficients with $\hat{\bm{\beta}}_1^{(TR-WEE)}$ and keep  $\frac{P_{R_i}({\bm{\hat{\gamma}}})-R_i}{1-P_{R_i}({\bm{\hat{\gamma}}})}E[ \bm{Q_{i1}}(Y_i,A_i,\bm{X_i;\hat{\alpha}^{(EE)},\hat{\beta}^{(EE)}}_1 )   |\bm{X,Y;\hat{\delta}}]$ as the augmented term, because we want to construct more robust estimators to protect against the misspecifications of either the missingness or the imputation model.

\section{Asymptotic Variance and Distribution for $\hat{\tau}$}
\label{sec_tr_dist}

The previous sections have discussed the point estimate for $\tau$. However, to obtain valid statistical inference, we also need the asymptotic variance and distribution of $\hat{\tau}$. This section discusses how to obtain the asymptotic distribution of $\hat{\tau}$ based on the Taylor expansion, delta method, and Slutsky's theorem. In this section, we do not distinguish the different methods to estimate $\tau$ (i.e. regardless of either IPW or TR method).

In Section \ref{sec_IPW} and \ref{sec_tr}, we have shown that  $E[\hat{\tau}_1]=\tau_1$, $E[\hat{\tau}_0]=\tau_0$ and they follow asymptotically normal under some ``conditions" depending on which method we use. For example, if we apply IPW-IPW method, we require both the missingness and PS models to be correct. If we use IPW-DR or IPW-WEE, we require the missingness model requires to be correct and another correct model from either PS or the outcome model. If we utilize TR-AIPW or TR-WEE, we require ``two out of three groups of models to be correct". 

In the first step, we need to obtain the joint distribution and covariance matrix for  $(\hat{\tau}_1,\hat{\tau}_0)$. Since the expectations of joint estimating equations for $(\hat{\tau}_1,\hat{\tau}_0)$ are zero, based on the properties of estimating equations, the joint distribution of $(\hat{\tau}_1,\hat{\tau}_0)$ will approximately follow a multivariate normal distribution (MVN), written as:
\begin{equation}
\begin{aligned}
\label{joint_dist}
\begin{pmatrix}
\hat{\tau}_1 \\
\hat{\tau}_0 
\end{pmatrix}	
\dotsim \text{MVN} 
\left(\begin{pmatrix}
\tau_1 \\
\tau_0
\end{pmatrix}
,\Sigma_{(2\times2)} \right)
\end{aligned}    
\end{equation}
where $\Sigma$ is $2 \times 2$ covariance matrix of $(\hat{\tau}_1,\hat{\tau}_0)$. The estimated covariance matrix $\hat{\Sigma}$ include the variance and covariance of $(\hat{\tau}_1,\hat{\tau}_0)$, denoted as $\widehat{Var}(\hat{\tau}_1),\widehat{Var}(\hat{\tau}_0),\widehat{\text{cov}}(\hat{\tau}_1,\hat{\tau}_0)$, which can be obtained based on the sandwich formula \cite{huber2004robust,freedman2006so}) or bootstrap approach \cite{davison1997bootstrap}.  

In the second step, we want to show that $\log(\hat{\tau})=\log(\frac{\hat{\tau}_1}{1-\hat{\tau}_1})-\log(\frac{\hat{\tau}_0}{1-\hat{\tau}_0})$ follows approximately normal distribution. The reason to make a log transformation is because of the non-linear form within $\hat{\tau}$ as the odds ratio. We denote $f=f(\tau_0,\tau_1)=\log(\tau)$ and we conduct two-dimension Taylor expansion for $\log(\hat{\tau})$ at true values $(\tau_0,\tau_1)$, written as
\begin{equation}
\begin{aligned}
\label{taylor_logtau}
\log(\hat{\tau}) 
&=f+\frac{\partial f}{\partial \tau_1} (\hat{\tau}_1-\tau_1)+\frac{\partial f}{\partial \tau_0} (\hat{\tau}_0-\tau_0)\\
& \indent +\frac{1}{2}\left[\frac{\partial^2 f}{\partial \tau_1^2} (\hat{\tau}_1-\tau_1)^2 +2\frac{\partial^2 f}{\partial \tau_1\partial \tau_0} (\hat{\tau}_1-\tau_1)(\hat{\tau}_0-\tau_0) +\frac{\partial^2 f}{\partial \tau_0^2} (\hat{\tau}_0-\tau_0)^2  \right]+\dots \\
&=\log(\frac{\tau_1}{1-\tau_1})-\log(\frac{\tau_0}{1-\tau_0})+\frac{1}{\tau_1(1-\tau_1)}(\hat{\tau}_1-\tau_1)+\frac{1}{\tau_0(1-\tau_0)}(\hat{\tau}_0-\tau_0) +R
\end{aligned}    
\end{equation}
where $R$ refers to the remaining terms including the second and higher orders of Taylor expansion. $R$ can usually be ignored because we know $\sqrt{n} (\hat{\tau}_1-\tau_1) \xrightarrow[]{\text{d}} N(0,Var(\hat{\tau}_1))$, which is $O_p(1)$, so  $\hat{\tau}_1-\tau_1=O_p(\frac{1}{\sqrt{n}})$. Similarly, we know $\hat{\tau}_0-\tau_0=O_p(\frac{1}{\sqrt{n}})$. Then, the remaining terms are not larger than $O_p(\frac{1}{n})$, which does not affect the variance as $n \to \infty$.

Due to the joint distribution of $(\hat{\tau}_1,\hat{\tau}_0)$  follows MVN from Equation (\ref{joint_dist}), we know that the linear combination of $(\hat{\tau}_1,\hat{\tau}_0)$ from Equation (\ref{taylor_logtau}) will also approximately follow the normal distribution, written as:
\begin{equation}
\begin{aligned}
\label{dist_logtau}
\log(\hat{\tau}) \dotsim N(\log(\tau),Var(\log(\hat{\tau})))
\end{aligned}    
\end{equation}

To estimate the asymptotic variance, after taking variance on both sides of Equation (\ref{taylor_logtau}) and plugging in estimated values of $(\hat{\tau}_1,\hat{\tau}_0)$, we will have:
\begin{equation}
\begin{aligned}
\label{var_logtau}
\widehat{Var}(\log(\hat{\tau}))=\widehat{Var}(\hat{\tau}_1)\frac{1}{[\hat{\tau}_1(1-\hat{\tau}_1)]^2}+\widehat{Var}(\hat{\tau}_0)\frac{1}{[\hat{\tau}_0(1-\hat{\tau}_0)]^2}+\widehat{\text{cov}}(\hat{\tau}_1,\hat{\tau}_0)\frac{1}{\hat{\tau}_1(1-\hat{\tau}_1)}\frac{1}{\hat{\tau}_0(1-\hat{\tau}_0)}
\end{aligned}    
\end{equation}
where $\widehat{Var}(\hat{\tau}_1),\widehat{Var}(\hat{\tau}_0),\widehat{\text{cov}}(\hat{\tau}_1,\hat{\tau}_0)$ are estimated variance and covariance from $\hat{\Sigma}$ in Equation (\ref{joint_dist}), respectively. 

In the last step, using the delta method, we can transfer $\log(\hat{\tau})$ back to $\hat{\tau}$. Since $\hat{\tau}=\text{exp}(\log(\hat{\tau}))$, based on results in Equation (\ref{dist_logtau}-\ref{var_logtau}), the asymptotic distribution and variance for $\hat{\tau}$ will be:
\begin{equation}
\begin{aligned}
\label{dist_tau}
\hat{\tau} \dotsim N(\tau,\tau^2 Var(\log(\hat{\tau})))
\end{aligned}    
\end{equation}
The estimated asymptotic variance is $\widehat{Var}(\hat{\tau})=\hat{\tau}^2 \widehat{Var}(\log(\hat{\tau}))$ and the 95\% confidence interval can be approximated by $\hat{\tau} \pm 1.96 \sqrt{\widehat{Var}(\hat{\tau})}$. As we described before, $\widehat{Var}(\hat{\tau}_1),\widehat{Var}(\hat{\tau}_0),\widehat{\text{cov}}(\hat{\tau}_1,\hat{\tau}_0)$ can be obtained based on the sandwich formula \cite{huber2004robust,freedman2006so}, but the explicit form is very complex because the joint estimating equations contains many equations and unknown parameters $(\gamma,\delta,\alpha,\beta,\tau_0,\tau_1,\tau)$. Instead, in both simulation and application studies, we can directly apply bootstrap approach to estimate the standard error $\sqrt{\widehat{Var}(\hat{\tau})}$.

As an alternative example, if we consider the continuous outcome and define $\tau=\tau_1-\tau_0$ as the main interest, the asymptotic normality for $\tau$ will be straightforward without using logarithm transformation and Taylor expansion, due to the simple linear form.


\section{True Causal Effect in the Simulation Study}

After the model is specified, the true causal effect needs to be found to calculate the bias. In this study, the true causal effect is defined as the odds ratio:
\begin{equation}
\begin{aligned}
\label{truecausal}
\tau=\frac{P(Y^1=1)/P(Y^1=0)}{P(Y^0=1)/P(Y^0=0)}
\end{aligned}
\end{equation}
We want to find the connection between the potential outcome and the model. Since the distribution of three covariates follows $N(0,1)$ independently in the simulation studies, $P(Y^1=1)$ can be simplified as:
\begin{equation}
\begin{aligned}
P(Y^1=1)
=& E[E(Y^1|A,\bm{X}_i))]  
= E[P(Y=1|A=1,\bm{X}_i)] 
= E \left[\frac{\text{\text{exp}}(\bm{X^T_i} \bm{\beta}+\beta_A)}{1+\text{\text{exp}}(\bm{X^T_i} \bm{\beta}+\beta_A)} \right] \\
=& \int_{-\infty}^{+\infty}\int_{-\infty}^{+\infty}\int_{-\infty}^{+\infty} \left[\frac{1}{\sqrt{2\pi}} \right]^3 \text{\text{exp}}\left[-\frac{x_1^2}{2}-\frac{x_2^2}{2}-\frac{x_3^2}{2} \right] \frac{\text{\text{exp}}(\bm{x^T_i} \bm{\beta}+\beta_A)}{1+\text{\text{exp}}(\bm{x^T_i} \bm{\beta}+\beta_A)} dx_1dx_2dx_3. 
\end{aligned}
\end{equation}
The first equality is due to double expectation conditioning on $(A,\bm{X}_i)$. The second equality is due to SITA. The third equality is due to the outcome model and the final equality is because of the independence among three covariates in the simulation study. 

Similarly, $P(Y^0=1)$ can also be written as:
\begin{equation}
\begin{aligned}
P(Y^0=1)
= \int_{-\infty}^{+\infty}\int_{-\infty}^{+\infty}\int_{-\infty}^{+\infty} \left[\frac{1}{\sqrt{2\pi}} \right]^3 \text{\text{exp}}\left[-\frac{x_1^2}{2}-\frac{x_2^2}{2}-\frac{x_3^2}{2} \right]  \frac{\text{\text{exp}}(\bm{x^T_i} \bm{\beta})}{1+\text{\text{exp}}(\bm{x^T_i} \bm{\beta})} dx_1dx_2dx_3.
\end{aligned}
\end{equation}
Then, $P(Y^1=1)$ and $P(Y^0=1)$ can be found by taking triple integration. After that, the true causal effect can also be found numerically as $\tau=2.247$.




\end{document}